%% file: main.tex
\documentclass[aps,prc,twocolumn,superscriptaddress,nofootinbib,tightenlines,
preprintnumbers,showkeys,floatfix]{revtex4-2}

\usepackage[utf8]{inputenc}
\usepackage[english]{babel}
\usepackage{amssymb,amsthm,amsmath,amstext,amsbsy,amsopn,mathrsfs}
\usepackage{bbm}
\usepackage{nicefrac}
\usepackage{slashed}
\usepackage{graphicx}
\usepackage[hidelinks]{hyperref}
\hypersetup{pdftitle={Geometric factors in the volume dependence of many-body bound states},pdfauthor={Hang Yu}}
\usepackage{leftidx}
\usepackage{environ}
\usepackage{mathtools}
\usepackage{xspace}
\usepackage{array}
\usepackage{xcolor}
\usepackage{pgfplotstable}
\usepackage{isotope}
\usepackage{suffix}
\usepackage{booktabs,colortbl}
\usepackage{physics}
\usepackage{braket}
\usepackage{orcidlink}
\input{defs.tex}

\renewcommand{\vec}[1]{\boldsymbol{#1}}

\begin{document}

\title{Geometric factors in the volume dependence of many-body bound states}

\author{Hang Yu\,\orcidlink{0000-0001-6860-5960}}
\email{yhang@nucl.ph.tsukuba.ac.jp}
\affiliation{Center for Computational Sciences, University of Tsukuba, Tsukuba, Ibaraki 305-3577, Japan}

\begin{abstract}
I derive the volume dependence of bound states from a cluster-cluster picture with nucleon degrees of freedom. A constant factor called the geometric factor appears in the generalization from pointlike particles to clusters. I show that this factor is obtained explicitly when the underlying overlap integral is evaluated directly in the original many-body space. I achieve this by constructing Jacobi coordinates on the lattice with periodic boundary conditions. The derivation requires only the interaction between the clusters to be short ranged, with the nonperturbative Coulomb force included. The factor emerges as a geometric property of the many-body configuration space rather than a multiplicity of physical channels. I validate my derivation using many-body calculations; in particular, I find this factor to be essential in extracting asymptotic normalization constants from lattice calculations of the \isotope[16]{O} ground state.
\end{abstract}

\maketitle

\section{Introduction}
Finite volumes and truncated bases are unavoidable in numerical few- and many-body calculations. The resulting truncation dependence of bound-state energies must therefore be controlled when extrapolating to the infinite-volume or infinite-basis limit. On periodic lattices, Lüscher's framework~\cite{Luscher:1985dn,Luscher:1986pf,Luscher:1990ux,Luscher:1991cf} has been extended to alternative boundary conditions~\cite{Briceno:2013hya,Bour:2011ef}, long-range interactions~\cite{Bubna:2024izx}, Coulomb interactions~\cite{Beane:2014qha,Guo:2021lhz,Guo:2021qfu,Yu:2022nzm}, and systems with more particles~\cite{Polejaeva:2012ut,Briceno:2012rv,Mai:2017bge,Doring:2018xxx,Hammer:2017kms,Meissner:2014dea,Briceno:2019muc,Konig:2017krd}. Related trapped-system and infrared-extrapolation relations play an analogous role for harmonic-oscillator (HO) bases~\cite{Busch:1998cey,Luu:2010hw,Zhang:2020rhz,Furnstahl:2013vda,More:2013rma,Furnstahl:2014hca}. These relations guide extrapolations and provide convergence tests in \textit{ab initio} calculations of finite nuclei and hadrons~\cite{BMW:2014pzb,Elhatisari:2025fyu,Miyagi:2023zvv}. In this paper, I argue that applying a pointlike formula to emergent clusters in a nucleon-resolved calculation can, however, omit a multiplicity carried by the many-body configuration space.

For a localized $A$-body state with an isolated leading breakup component $A\to(A-C)+C$, I show that this multiplicity appears as a constant $G_{A,C}$ multiplying the standard point-particle energy shift. That shift already contains the ordinary factor counting spatial directions. I find that this spatial-direction factor and the new factor come from the same source: both count leading-order faces of the many-body domain, the former through the spatial directions and the latter through the coordinate-labeled faces with a fixed composition of conserved labels. Therefore, I call $G_{A,C}$ the geometric factor. At leading order, $G_{A,C}$ and the asymptotic normalization constant (ANC) $C_0$ enter the energy shift only through the product $G_{A,C}C_0^2$, where $C_0$ is defined by the cluster-relative wave function with the intrinsic fragment states normalized to unity. The factor is therefore not separately observable: its meaning is tied to the microscopic representation and to this clustered convention. Omitting $G_{A,C}$ within that convention rescales an ANC inferred from the volume dependence. ANCs describe the normalization of asymptotic wave function tails and determine peripheral reaction observables~\cite{deBoer:2017ldl}. They are thus observables directly accessible to many numerical simulations. They also encode asymptotic cluster amplitudes with reduced sensitivity to details of the nuclear interior~\cite{Freer:2017gip}. Such observables are particularly valuable for charged-particle reactions, whose astrophysically relevant energies are difficult to access experimentally. Evaluating these observables hence becomes an important task for reaction and continuum algorithms~\cite{Nollett:2011qf,Yaron:2022rmb,Yu:2022nzm,Harris:2025zsh}. These geometric factors bridge the gap between Lüscher-type formulation and actual nucleosynthesis calculations.

Reference~\cite{Konig:2017krd} observed and hypothesized the bound-state volume dependence as a sum over breakup channels and identified a combinatorial factor associated with partitions of particles. That observation motivates the present work, but it does not consistently predict the counting rule for a nucleon-resolved cluster calculation. First, this channel cannot be physical. Permuting identical nucleons within a fixed cluster component does not generate distinct physical breakup channels, so counting physical channels cannot determine the required factor. A physical channel-counting explanation would also leave signatures in experimental observables, while two-body descriptions (for example, halo EFTs~\cite{Hammer:2017tjm}) are immune to such a signature. Second, the number of possible partitions is not unique---which microscopic labels used in many-body methods define equivalent partitions? For example, should one include quark rearrangements via pion exchange? My method gives a consistent answer to these questions.
My derivation predicts that tensor and spin-orbit forces will change $G_{A,C}$ when they mix individual spin projections, even though the physical cluster component is unchanged. For example, in the case of $\alpha$ clusters within \isotope[16]{O}, I predict that the geometric factor changes from $256$ in the spin-resolved counting (for example, SU(4) symmetric interaction~\cite{Lu:2018bat}) to $784$ when only the proton and neutron numbers remain conserved (for example, the conventional chiral interactions~\cite{Entem:2003ft,Machleidt2011,Ekstrom:2015rta,Ekstrom:2017koy}).

I fix the counting rule by evaluating the periodic-image overlap~\cite{Yu:2022nzm,Yu:2023ucq} directly in the original many-body space. The technical ingredient is a lattice-compatible Jacobi domain: the Jacobi linear combinations are conventional, but periodic translations are used to construct a box-shaped fundamental domain with controlled boundary identifications. This construction preserves the cluster-relative coordinate and removes the center-of-mass ambiguity that otherwise arises under repeated boundary shifts~\cite{Konig:2022cya}. It also allows Green's identity to reduce the many-body overlap to a manageable cluster-relative surface term while retaining the Coulomb-distorted tail. Compared to the previous work, the following contributions are worth noting: (i)~I compute the actual many-body overlap integral, making the geometric factor and its counting rule explicit. (ii)~The assumptions are weaker. Only the non-Coulomb interaction between the clusters needs to be short ranged---for the leading image translations that displace a cluster as a whole, its internal separations are unchanged and all intracluster interactions cancel from the translated potential difference---and the intercluster Coulomb force is retained nonperturbatively. By contrast, Ref.~\cite{Konig:2017krd} assumes every interaction to vanish beyond a strictly finite range. (iii)~The factor emerges as a geometric property of the many-body configuration space, independent of particle statistics and determined by the conserved quantum labels, rather than as a multiplicity of physical channels.

Throughout the derivation, I assume that the fragments are localized and small compared with the box, and that the leading two-cluster component is well separated from subleading thresholds; beyond two-cluster shifts are exponentially suppressed by this localization. If several components have comparable asymptotic scales, the energy shift becomes a sum of their leading contributions, each with its own overlap normalization and geometric factor. The closed Whittaker-squared form used below applies to a three-dimensional (3D) cubic box and $\ell=0$; higher partial waves require numerical evaluation of the surface integral.

I test the counting rule in both HO and periodic settings. Calculations of \isotope[4]{He} with the in-medium similarity renormalization group (IMSRG) and of \isotope[4]{He} and \isotope[20]{Ne} with nuclear lattice effective field theory (NLEFT) provide clean numerical checks. The \isotope[16]{O} example is a unique test: in the accessible volumes, the large geometric factor amplifies the subleading $\alpha$ component enough to change the sign and apparent scale of the net volume dependence; I discuss the resulting fit-window sensitivity in Sec.~\ref{sec:numerics}. Section~\ref{sec:mean_field} first exposes the geometric origin of the factor in a simple Dirichlet-wall model relevant to HO truncations. Section~\ref{sec:Jacobi_Lattice} constructs the periodic Jacobi domain, Sec.~\ref{sec:main} derives the many-body counting rule, Sec.~\ref{sec:scattering} briefly discusses the scattering region, and Sec.~\ref{sec:numerics} presents the numerical tests and their limitations.

\bigskip
\section{A simple case with Dirichlet boundary}
\label{sec:mean_field}
{I begin by deriving the geometric factor for a simple case with Dirichlet boundary conditions, before introducing Jacobi lattices.}
{In practice, Dirichlet boundaries usually arise in calculations using} harmonic oscillator (HO) bases. 
HO bases are dominant in many different single-particle basis models~\cite{Hergert:2020bxy,Hagen:2013nca} and no-core shell models~\cite{Barrett:2013nh}.
The exponential cutoff dependence~\cite{Furnstahl:2013vda,More:2013rma}
\begin{equation}
    \Delta E_{\textrm{HO},L} = \frac{\kappa C_0^2}{\mu}\exp(-2 \kappa L)
\end{equation}
has been one of the most commonly used tools to check the consistency and convergence of these calculations, with the constant multiplier often referred to as related to the ANCs. Here $\Delta E_{\textrm{HO},L} = E_L - E_\infty$ denotes the shift of the bound-state energy at an effective basis size $L$, $\kappa$, and $\mu$ are the binding momentum and the reduced mass of the lowest breakup channel, and $C_0$ is the corresponding ANC.
Yet, the true relation is rarely derived and implemented in the calculations, while the cutoff dependence can be huge in heavy nuclei.
The large sizes of the heavy nuclei are often debated to be the cause of strong IR cutoff dependencies. 
I will show that this factor can also be geometrically increasing, simply owing to the large configuration space (in Sec.~\ref{sec:main}). 
With ANCs constrained by the normalization, the volume dependence and IR cutoff is amplified simply by the choice of nucleons as the intrinsic degrees of freedom as opposed to clusters. 
It is possible that this amplification marks an urgent need for precision formulas to remove this volume dependence and IR cutoff in these \textit{ab initio} calculations, reducing the systematic errors and improving the understanding regarding heavy elements.

The system considered throughout this paper is
\begin{multline}
        H \mathcal A= \left(-\sum_{i=1}^A \frac{\nabla_i^2}{2m} + V(\vec r_1,\vec r_2,\ldots,\vec r_A) \right)\mathcal A 
        \\ = E_\infty \mathcal A
    \label{eq:hamilton}
\end{multline}
with the assumption that the basis is large enough to represent the short-range interaction $V$, hence the details of $V$ are  not important. Here $\mathcal A$ denotes the $A$-body bound-state wave function with energy $E_\infty$, $\vec r_i$ are the single-particle coordinates, and $m$ is the nucleon mass.

In finite volumes, each single-particle coordinate is restricted to a region $S$ of half-width $L$. In the 1D discussion below, $S = [-L,L]$, and the many-body domain is the hypercube $S^A$. I write
\begin{multline}
    H_L \mathcal A_L = \left(-\sum_{i=1}^A \frac{\nabla_i^2}{2m} + V(\vec r_1,\vec r_2,\ldots,\vec r_A) \right)\mathcal A_L
    \\ = E_L \mathcal A_L\,.
    \label{eq:hamilton-truncated}
\end{multline}
The difference is in the boundary condition: I require $\mathcal A_L(\vec r) = 0$ at $r_i = L$, replacing the condition $\mathcal A(\vec r) \to 0$ as $r_i \to \infty$. The reason I distinguish $H$ and $H_L$ is that in practice, the boundary conditions are imposed via a specific choice of basis truncation, resulting in different representations of the operator. The aim of Lüscher's formalism (for bound states) is to obtain a well-defined description of the energy shift $\Delta E_L =E_L  -  E_\infty $.
Several methods exist to derive the volume dependence for the Dirichlet boundary, which can later serve as an approximation to HO basis truncation. 
Of all these methods, it is necessary to guarantee that the boundary condition is satisfied and the shift to the bound-state energy emerges:
\begin{multline}
    \psi_{L}(r)\approx C_0 [\exp(-\kappa r) -\exp(-2\kappa L) \exp(\kappa r) ]\\
    \approx \psi(r)-\psi(2L-r)\,,
\end{multline}
for $r \to L$. 
I start my derivation for the factor with the Dirichlet boundary based on this ansatz. 

First, consider the single-particle basis, widely used in many-body methods for heavy nuclei such as IMSRG and coupled-cluster. These methods start from reference states filling a set of single-particle energies, with the bulk energy the sum of the noninteracting energy of each particle. The antisymmetric construction of Hartree-Fock states generally will not change the bound-state volume dependence~\cite{Luscher:1985dn,Furnstahl:2013vda}, and I consider the simplest example, scalar particles with the same mass and in the same mean-field trap  $ V_{\mathrm{SP}}$ to avoid symbolic complications. From this basic construction, I write the equation in one dimension and in the one-body potential case:
\begin{equation}
        H = \sum_{i=1}^A \left(-\frac{\nabla_i^2}{2m} + V_{\mathrm{SP}}( x_i)\right) \equiv \sum_{i=1}^A H_{\mathrm{SP}}(x_i)\,.
    \label{eq:ni-hamilton}
\end{equation}
I denote $ E_\infty= A\cdot E_{\mathrm{SP}}$.
Similarly, $H_L =\sum_{i=1}^A H_{\mathrm{SP},L}(x_i) $, which supports a bound state $\psi_L(x_i)$ for each index $i$.\footnote{
I note that the choice of $x_i$ over radial $r_i$ here is intentional: the infinite $\psi$ cannot be fully represented by truncated basis, and the truncated $\nabla^2$ is hence not self-adjoint on this domain if applied to $\psi$. Choosing $r_i$ will cause both inconvenience (to shift by $L$) and confusion (on the boundaries) later.} 
In the infinite volume, the wave function is a product state:
\begin{equation}
    \mathcal{A} (\vec x)  = \prod_{ i} \psi(x_{i})  \,,
\end{equation}
and $\psi(x_i) \to \frac{C_0}{\sqrt{2}} \exp(-\kappa x_i)$ when $x_i \to \infty$. The $1/\sqrt{2}$ factor comes from the switch to $x_i$ from the radial coordinate.
I then {consider} the following {ansatz in} the $A$-body picture:
\begin{multline}
    \mathcal{A}_{L} (\vec x) \equiv \prod_{ i}  \psi_{L}(x_i) \\
    = \underbrace{\mathcal A(\vec x) 
    - \sum_{i=1}^A \left[\mathcal A(2\hat x_i L -\vec x )
    +\mathcal A(-2\hat x_i L -\vec x )\right]}_{A_{L,0}(\vec x)} \\
    + O[\exp(- 2\kappa L)]\,.
\label{eq:ansatz_MF}
\end{multline}
From here, in fact, one can trivially arrive at [with $\left[-\frac{\nabla^2_i}{2m} +  V_{\mathrm{SP}}( x_i)\right]\psi_L(x_i) = E_{\mathrm{SP},L}\psi_L(x_i) $]
\begin{multline}
    H_L \mathcal{A}_{L} (\vec x) =\sum_{i=1}^A H_{\mathrm{SP},L}(x_i)\psi_L(x_i)\times  \prod_{ j\neq i }  \psi_{L}(x_j)  \\
    = A \times E_{\mathrm{SP},L}\mathcal{A}_{L} (\vec x)  \,.
\end{multline}
It is natural to find the energy shift $\Delta E_L = A \times( E_{\mathrm{SP},L} - E_{\mathrm{SP}}) = A \times \Delta E_{\mathrm{SP},L} $.
In this trivial picture, each single-particle state will acquire its own energy shift with respect to the boundary of basis, and as such, the volume correction will accumulate, resulting in a factor equal to the number of particles. Hence, this factor is easily mistaken for the number of all possible channels.
However, this simple explanation encounters conceptual difficulties in multinucleon clusters as nucleons are identical particles, where numerical evidence in Sec.~\ref{sec:numerics} suggests that such a factor always exists.

Therefore, to derive this factor in a consistent and general manner, I use the conventional overlap integral method, which first appeared in Lüscher's original bound state paper~\cite{Luscher:1985dn} and has been extended to the long-range Coulomb interaction~\cite{Yu:2022nzm}.
As discussed above, volume dependence emerges already at the single-particle level, and neither fermionic nor bosonic statistics affects this conclusion. I therefore focus on the same setup: scalar noninteracting particles on a finite basis and zero angular momenta (to avoid doing permutation in the notations).
I use this trivial noninteracting example to validate and explain my formulation.
This $A_{L,0}$ form defined in Eq.~\eqref{eq:ansatz_MF} satisfies the following conditions, making it a suitable candidate to approximate volume dependence:
\begin{enumerate}\renewcommand{\labelenumi}{(\roman{enumi})}
    \item $\mathcal{A}_{L,0} (\vec x) $ needs to vanish at the boundaries $x_i = \pm L$, \\
    \item $\lim_{L\to \infty} \mathcal{A}_{L,0} (\vec x) = \mathcal A (\vec x)$,\\
    \item $\mathcal{A}_{L,0} (\vec x) = \mathcal{A}_{L} (\vec x) +O(\exp(- 2\kappa L))$ on the domain $S^A$.
\end{enumerate}
The first condition guarantees the hard wall boundary condition, the second condition fixes the level under consideration when $L\to \infty$. The last condition ensures the desired level of approximation to study volume dependence.  The first two conditions are straightforward. To get the third condition in the Lüscher's formalism (i.e., using only the short-range properties of $V$), I define $\eta (\vec x)$ by:
\begin{equation}
    H_L\mathcal{A}_{L,0} =  E_\infty \mathcal{A}_{L,0} (\vec x) +\eta (\vec x)\,.
    \label{eq:diff-hw}
\end{equation}
Note that
\begin{multline}
  H_L   \mathcal A(2\hat x_i L -\vec x )= \\
   \left[H(2\hat x_i L-\vec x) + H(\vec x) - H(2\hat x_i L-\vec x) \right] \mathcal A(2\hat x_i L -\vec x )\\
  =  [E_\infty + V(\vec x) - V(2 \hat x_i L - \vec x) ]   \mathcal A(2\hat x_i L -\vec x )\,,
\end{multline}
where I use $ H(\vec x) = -\nabla^2/2m + V(\vec x)$ to note the subsequent shift on $\vec x$.
And $x_i\in[-L, L]$ from $H_L$,
\begin{multline}
   \eta (\vec x) \sim  ~
   2\sum_{j=1}^A \prod_{i\neq j} \psi(x_i)\\    [  V(2 \hat x_j L - \vec x) - V(\vec x)]
   \psi(2L- x_j) \\ \sim O[\exp(- 2 \kappa L)]\,.
   \label{eq:eta-hw}
\end{multline}
As a consequence, one can expect $\mathcal{A}_{L,0} =\mathcal{A}_{L} +O[\exp(-2\kappa L)]$, i.e., a good approximation of {the} true wave function at sufficiently large $L$ stated in the third criterion. I will omit exponentially suppressed terms from now on. I follow the approximation in~\cite{Luscher:1985dn} that:
\begin{multline}
    \Delta E_L \approx \braket{\mathcal A|\eta} =2 \sum_{j=1}^A\int_{S^{ A}} \prod_i\dd x_i \prod_i \psi(x_i) \\ \times [ H(2\hat x_j L- \vec x )-H(\vec x)] \psi(2L- x_j)\prod_{i\neq j} \psi(x_i)\,.
    \label{eq:1d-addzero}
\end{multline}
$S^A$ is the $A$-dimensional finite box. Note that $\psi(2L- x_j)\prod_{i\neq j} \psi(x_i)$  is an eigenvector of $H(2\hat x_j L- \vec x )$,
\begin{equation}
    \Delta E_L =2\sum_{j=1}^A \int_{S^A} \prod_i\dd x_i\, \mathcal A(\vec x )  [E_\infty- H] \mathcal A(2L \hat x_j - \vec x) \,.
    \label{eq:overlap-ho}
\end{equation}
I then {follow} the surface-integral technique with details in the supplemental material of Ref.~\cite{Yu:2022nzm}, because of the
\textit{non-Hermiticity} here: $\psi(x_i)$ and $\psi(2L- x_i)$ are not wave functions for which the kinetic operator $\nabla_i^2$ is self-adjoint on the domain $S$.
Using integration by parts, $u v'' =  (uv'-u'v)' + u''v$,
\begin{multline}
    \Delta E_L
    =- \frac{ 2A}{m}\psi'(L) \psi(L) \\
    +2\sum_{j=1}^A \int_{S^A} \prod_i\dd x_i\,  \mathcal A(2L \hat x_j - \vec x) [E_\infty- H]\mathcal A(\vec x ) \\
    = -\frac{2A}{m}\psi'(L) \psi(L)\,.
\label{eq:1dsurface}
\end{multline}
{In the noninteracting picture, the above result relates to the single-particle energy shift as:}
\begin{equation}
    \Delta E_L =A \Delta E_{\textrm{SP},L} =A \times \frac{\kappa C_0^2}{m}\exp(-2 \kappa L) \,.
\end{equation}
This simple noninteracting toy model generalizes well to the $(A{-}1)$--$1$ cluster configuration once one has the infinite-volume wave function $\mathcal A(\vec x)$ with nonscalar particles.
A similar ansatz can be used as in Eq.~\eqref{eq:ansatz_MF},
where
\begin{equation}
   \mathcal{A}_{L,0} (\vec x) \propto\mathcal A(\vec x) - \sum_{i=1}^A \left[  \mathcal A(2\hat x_i L -\vec x )+\mathcal A(-2\hat x_i L -\vec x )\right]\,.
\end{equation}
By assuming the center of mass to be at zero, I can factorize $\mathcal A(\vec x)$ into $\psi(x_i)\times \text{remaining}$ approximately around the boundary~\cite{Konig:2017krd}.
It is then easy to check that the three conditions proposed above are satisfied, 
and the rest of the derivation follows. In this more complicated case, the additional factor $A$ counts the number of faces of the hypercube in the $A$-dimensional configuration space for this $(A{-}1)$--$1$ cluster configuration, while the channel/multiplicity explanation cannot explain how the channels are counted physically versus numerically.
The mechanism to include how clusters emerge from the single-particle description is beyond the scope of this paper (and necessarily involves relative coordinates $\vec x_i - \vec x_j$), but for a general $(A{-}a)$--$a$ cluster configuration, a similar statement holds (see Sec.~\ref{sec:main} for the periodic boundary case): this factor equals the number of faces of the relevant polytope in configuration space. This factor also grows combinatorially with the cluster mass numbers. Hence, I refer to it as the geometric factor.  For more complex box geometries, such as a hexagonal prism, the factor will differ, and is in general independent of either the state or the multiplicity of channels.

\bigskip

\section{Jacobi coordinates}
\label{sec:Jacobi_Lattice}
On the lattice, periodic boundaries impose a torus topology---translational invariance can be visualized by identifying opposite faces of the hypercube.  Lüscher's original construction~\cite{Luscher:1985dn,Luscher:1986pf,Luscher:1990ux} is, in fact, formulated on a 3-torus. Here I generalize to an $A$-body picture, for which I need to construct (partial) Jacobi coordinates on hyper-tori.  This step is what allows me to evaluate the overlap integral directly in the original many-body space: the surface integrals generalizing Eq.~\eqref{eq:1dsurface}, which relax the assumptions on which interactions need to be short ranged, would otherwise become too complicated to represent symbolically in existing coordinate systems.

Jacobi coordinates can be difficult to use in lattice formulations. For example,  the lattice sites often create relative coordinates that do not coincide with existing lattice sites, and the center of mass can be multiply defined.
This center-of-mass ambiguity compounds the already exponentially scaling cost of few-body methods~\cite{Konig:2022cya}. Light systems can alternatively be solved variationally in the periodic box without constructing relative coordinates explicitly~\cite{Yaron:2022rmb}; such direct few-body treatments are complementary to the closed-form volume dependence derived here for clusterized nuclei. As a side product of my derivation of the geometric factor, I find a construction that elegantly avoids this ambiguity. I expect this construction to be independently useful for nuclear lattice calculations, and a recent study confirms that computing finite nuclei on lattices offers distinct advantages~\cite{Rothman:2025uza}. I therefore dedicate this section to the Jacobi lattice construction.

\begin{figure*}
    \centering
    \includegraphics[trim={20mm  45mm 15mm 5mm}, width=0.7\linewidth]{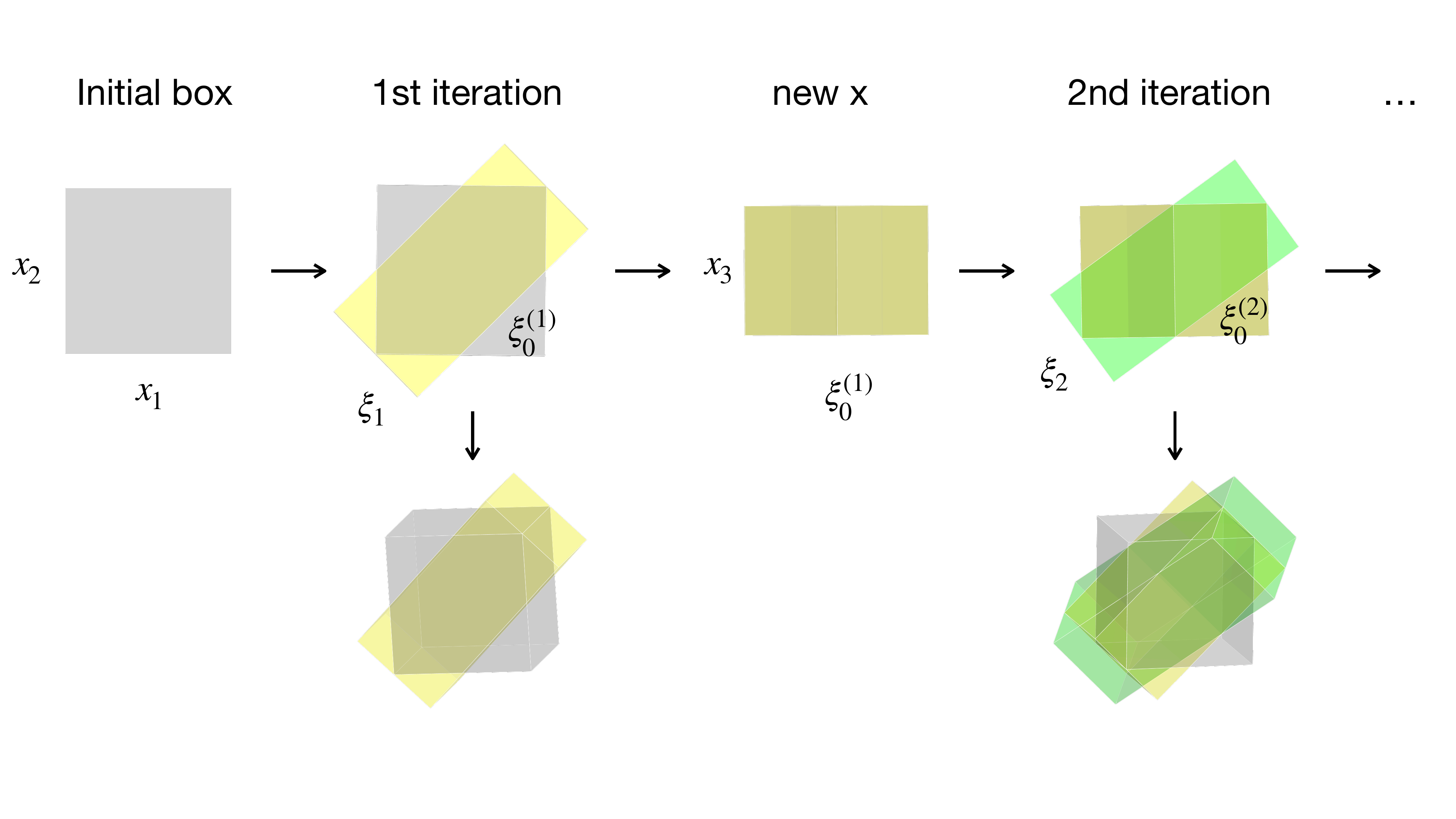}
    \caption{Iterative coordinate transformation from Cartesian to Jacobi lattice. Each subsequent step transforms one pair of coordinates $(\vec x_{i+1}, \vec \xi_0^{(i-1)}) \to (\vec \xi_i, \vec \xi_0^{(i)})$ while preserving the total volume and the simple boundaries. The top panels show the 2D coordinate plane at each iteration; the bottom panels show the corresponding 3D projection on the lattice.}
    \label{fig:ite}
\end{figure*}

\begin{figure}
    \centering
    \includegraphics[trim={20mm  5mm 15mm 5mm}, width=1.0\linewidth]{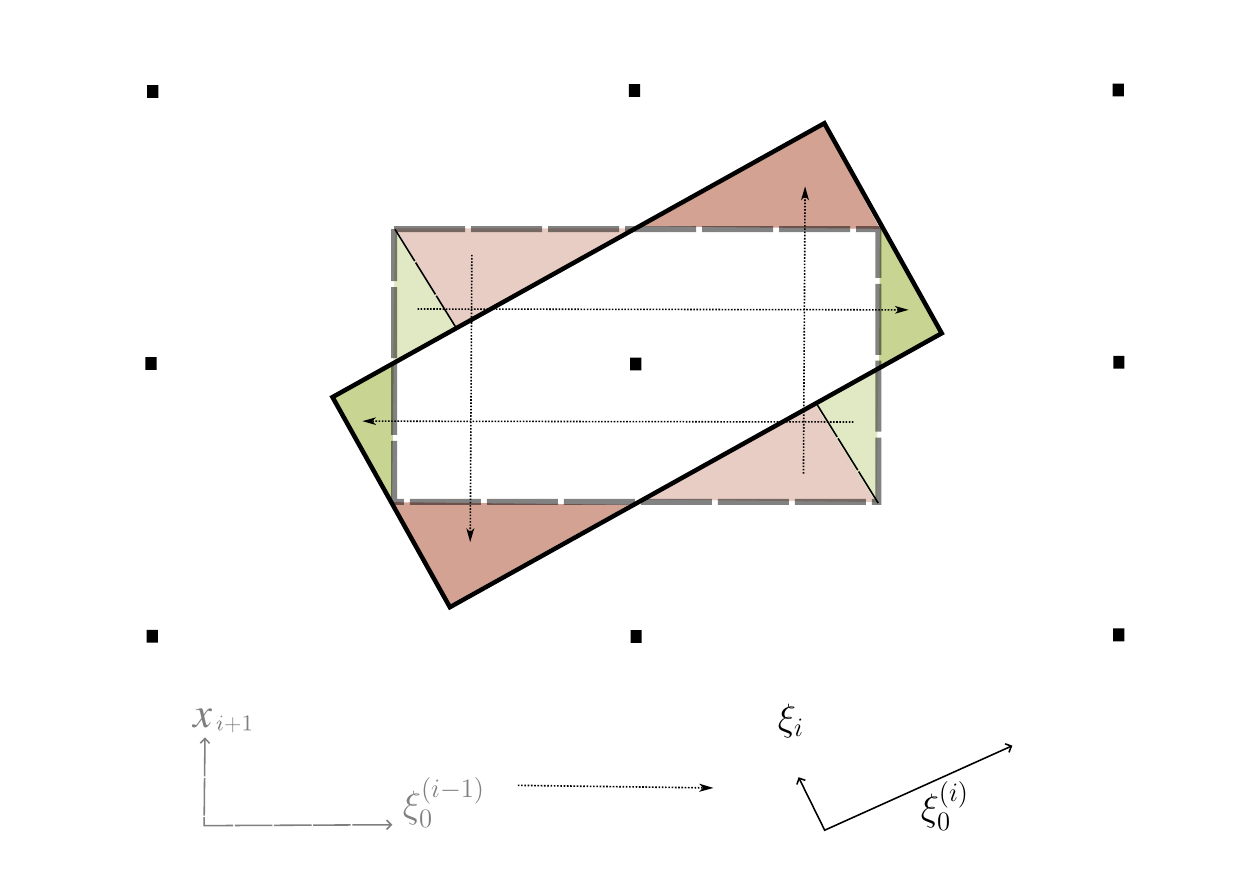}
    \caption{Single step of the coordinate transformation at the $i$th iteration. Transparent regions are shifted to opaque regions of the same color by $L$ (arrows), producing a rectangular region aligned with the Jacobi coordinates. This new region is equivalent to the original cubic lattice under the periodic boundary condition.}
    \label{fig:coord}
\end{figure}

The construction proceeds iteratively, as illustrated in Fig.~\ref{fig:ite}. Each step is depicted in Fig.~\ref{fig:coord}, and the procedure is as follows:
\begin{enumerate}\renewcommand{\labelenumi}{(\roman{enumi})}
    \item The aim is to describe cluster $C$ in a nucleus with $A$ particles on the cubic lattice with edge length $L$. To that end, I start with $d\times A$ dimensions of Cartesian coordinates, $\vec x_1,\vec x_2\ldots,\vec x_{A}$ and transform the first $d\times C$ coordinates  $\vec x_1,\vec x_2\ldots,\vec x_{C}$ to Jacobi coordinates.
    \item For the first iteration $i=1$, I find the center of mass (COM) for the first two coordinates, $\vec x_1,\vec x_2$. I write this COM as $\vec \xi_0^{(1)} =(\vec x_1 +\vec x_2)/2$. Then $\vec \xi_1 = \vec x_1 - \vec x_2$, perpendicular to $\vec \xi_0^{(1)}$. This is just the conventional Jacobi transformation. On the lattice I shift the cubic region according to Fig.~\ref{fig:coord} to the new Jacobi box using invariant translation operations on the torus. As a result, the volume remains unchanged, and the Jacobi coordinates remain also cubic, $(\vec \xi_1, \vec \xi_0^{(1)}) \in [-L/2,L/2]^d\times[-L/2,L/2]^d$ (cubes of side $L$).
    This first iteration and its projection to three dimensions are visualized in Fig.~\ref{fig:ite}.
    \item For each subsequent step $i>1$, I transform $\vec \xi_{0}^{(i-1)}$ with a new Cartesian coordinate $\vec x_{i+1}$  to Jacobi coordinates: $\vec \xi_i = \vec x_{i+1} - \vec \xi_{0}^{(i-1)}$, perpendicular to the center of mass $\vec \xi_0^{(i)} = (\vec x_{i+1}+ i  \vec \xi_{0}^{(i-1)})/(i+1)$.
    On the lattice I shift the cubic region according to Fig.~\ref{fig:coord} as in the first step, and $(\vec \xi_i, \vec \xi_0^{(i)}) \in [-L/2,L/2]^d\times[-L/2,L/2]^d$.
   The second iteration and its projection to three dimensions are visualized in Fig.~\ref{fig:ite}.
    \item The iteration terminates when the desired cluster center of mass is reached at  $i = C-1$, which produces partial Jacobi coordinates on the lattice with the remaining lattice coordinates untouched. Choosing $C= A$ yields the complete Jacobi coordinates including the center of mass for the whole system. Alternatively, I can transform the remaining $A-C$ particles into the Jacobi lattices and thereby define a total COM position. I can then offset the total COM to the origin and factor out this degree of freedom.
\end{enumerate}

This procedure rewrites the conventional lattice Cartesian coordinates with (hyper)cubic boundaries into Jacobi coordinates with (hyper)cubic boundaries. One can expect the surface integrals to be clearly defined on rectangular regions, and easy to represent symbolically after properly identifying the cluster composition. This allows, in this paper, easy generalization to arbitrary hyper-tori without additional assumptions and  unnecessary complications. A direct evaluation in the hyper-tori space guarantees all many-body contributions in the desired degree of freedom are properly included.

One can verify that these transformations remove the center-of-mass ambiguity:
each iteration in Fig.~\ref{fig:coord} eliminates one pair of COM ambiguity caused by shifting an axis by $L$. As a result, the COM $\xi_0$ on the lattice is uniquely defined modulo $L$.

This transformation does come at a cost: while periodicity remains untouched along the COM axis $\vec \xi^{(i)}_0$,
the boundary condition on $\vec \xi_i$ is more involved.  $\vec \xi_i + L$ still maps back to $\vec \xi_i $, but with an additional offset of $\vec \xi^{(i)}_0$ by $ \frac{L}{i+1}$. In Fig.~\ref{fig:coord}, when a particle moves out of the long side of the Jacobi lattice, it reenters on the other side with the same color. In practice, this modified boundary condition can be straightforwardly handled by returning to the Cartesian lattice for the indexing. Since this construction involves only iteratively shifting copies of triangles under the periodic boundary condition, the kinematics remains invariant. Computationally, the additional cost would amount to reordering indices to compute interactions in the relative basis.

To summarize, I have constructed an iterative procedure that maps Jacobi coordinates to a box-shaped region on the lattice. This enables rigorous evaluation of the integrals needed to extract the leading-order volume dependence for bound states in cluster configurations. I emphasize that the final counting rule derived in Sec.~\ref{sec:main} does not depend on the detailed choice of the internal Jacobi coordinates, provided the fragments are localized and small compared with the box. The role of the construction is to supply a well-defined torus domain on which the many-body overlap integral can be converted into the cluster surface integral. It does so without losing the internal coordinates or introducing an ambiguous center of mass. I expect this construction to be valuable for locating clusters on the lattice and for optimizing few-body algorithms such as the DVR method~\cite{Konig:2020lzo}.

\bigskip
\section{Geometric factors on hyper-tori}
\label{sec:main}
In this section, I aim to extend the previously derived volume dependence for a two-body system
\begin{equation}
 \Delta E_{\textrm{TB},L} = %
  {-}\frac{3\anc^2}{\mu L}
  \Big[W_{-\bar\eta,\frac{1}{2}}(\kappa L)\Big]^2
 +\ldots\,
\label{eq:DeltaE-3D-final}
\end{equation}
to the cluster-cluster configuration.
Here $\mu$ is the reduced mass, $\kappa$ the binding momentum of the system, $L$ the size of the cubic box, $\bar\eta = \frac{\mu Z_1 Z_2}{\kappa}$ the Sommerfeld parameter that describes the Coulomb interaction strength, and $W_{-\bar\eta,\frac{1}{2}}$ the Whittaker function. Importantly, $C_0$ is the asymptotic normalization constant for angular momentum $l=0$. For more details, I refer to Ref.~\cite{Yu:2022nzm}.
This formula describes the leading volume dependence of the bound state energy when Coulomb interaction is present in a periodic box, and it takes a simple squared form in the special case of angular momentum $l = 0$. Generalization to higher angular momenta requires numerical evaluation of a surface integral. The important distinction of this simple formula lies in the nonperturbative Coulomb force, which reduces to the exponential form when $\bar\eta\to 0$. The nonperturbative Coulomb interaction allows exploration of arbitrarily large $\bar \eta$, for example, in the case of shallow binding where $\kappa$ is small.

This formula can be readily generalized to clusters with the help of \citet{Konig:2017krd}. However, there is one question not clearly stated in that work: the computation of the so-called combinatorial factors, especially for cluster-cluster configurations. To explicitly derive this factor, which I call the geometric factor, I consider a similar setup as in Sec.~\ref{sec:mean_field}, with the help of Jacobi lattices derived in Sec.~\ref{sec:Jacobi_Lattice}.

For clarity, I collect the assumptions used throughout this section: (i) the $A$-body ground state is localized, with a dominant two-cluster breakup channel $A \to (A-C)+C$ characterized by the binding momentum $\kappa$, well separated from subleading channels characterized by $\kappa^*$; (ii) the intrinsic sizes of the two fragments are small compared to the box size $L$; (iii) the interaction between the fragments is short-ranged apart from the Coulomb force, while the interactions inside each fragment are arbitrary; (iv) the single-nucleon spin and isospin projections are conserved (this will be relaxed at the end of this section). I emphasize that no confined-range assumption on the individual potentials is needed.

The first step is to make an ansatz of a finite-volume wave function from the infinite volume.
As a heuristic motivation for the periodic structure, consider a trapped $A$-body wave function in a finite volume with size $L$, subject to a one-body trap $V$. As discussed above, the assumption of a one-body potential does not affect the objective of deriving this constant factor.
I expect the previous ansatz for the noninteracting picture Eq.~\eqref{eq:ansatz_MF} in Sec.~\ref{sec:mean_field}, which relates the many-body finite-volume ansatz to the one-body ansatz, to take the form
\begin{equation}
    \ket{\mathcal{A}_{L}}\equiv \ket{\psi_{1,L}, \psi_{2,L},\ldots, \psi_{A,L}} = \prod_i\ket{\psi_{i,L}}\,,
\end{equation}
with now $\psi_{i,L}(\vec x_i) \approx \sum_{\vec n_i}\psi_i(\vec x_i+\vec n_i L)$ following from existing methods~\cite{Luscher:1985dn,Konig:2017krd}. A quick estimate can be made using this product ansatz, and compared to the two-body pointlike case, this ansatz contains additional terms:
\begin{multline}
    \Delta E_L =
    \sum_{ \vec n_j, j } \int_{B^{d\times A}} \prod_i\dd \vec x_i  \\ \underbrace{\prod_i \psi_i(\vec x_i)}_{\mathcal{A} (\vec x )}
    \times  V(\vec x)\underbrace{\psi_j(\vec x_j+ \vec n_j L)\prod_{i\neq j} \psi_i(\vec x_i)}_{\mathcal{A} (\vec x +\vec n_j L )}+ \dots\,,
\end{multline}
where $\vec n_j$ denotes a shift on the component of the $j$th Cartesian coordinate and $\dots$ contains terms shifting two or more one-body factors, which are further exponentially suppressed (in this one-body trap case).
In the interacting picture, it is then natural to write the ansatz as (due to emergent clusters, shifts by multiple $\vec n_i$ are possible):
\begin{equation}
    \mathcal{A}_{L,0} (\vec x) = \sum_{\vec n = (\vec n_1,\vec n_2,\ldots,\vec n_{A})\in \ZZ^{d\times A}}   \mathcal{A} (\vec x + \vec n L)\,.
\label{eq:ansatz_lattice}
\end{equation}
For convenience, the Cartesian coordinates are denoted by $\vec x = (\vec x_1, \vec x_2,\ldots,$ $ \vec x_A)$, with $x_i \in \mathbb{R}^d$. Note I restrict $\vec x$ to $(B\equiv [-L/2,L/2])^{d\times A} $ on the lattice. I also assume (without loss of any generality) the origin to be the center of mass in $\mathcal{A}(\vec x)$, and shifting all $\vec x$ together by multiples of $L$ will always be the vanishing tails.
This form satisfies the analogous conditions in Sec.~\ref{sec:mean_field}, revised for the periodic boundary:
\begin{enumerate}\renewcommand{\labelenumi}{(\roman{enumi})}
    \item $\mathcal{A}_{L,0} (\vec x) $ is periodic: $\mathcal{A}_{L,0} (\vec x + \vec e_i L) = \mathcal{A}_{L,0} (\vec x) $, \\
    \item $\lim_{L\to \infty} \mathcal{A}_{L,0} (\vec x) = \mathcal A (\vec x)$,\\
    \item $\mathcal{A}_{L,0} (\vec x) = \mathcal{A}_{L} (\vec x) +O[\exp(- \kappa L)]$.
\end{enumerate}
where the momentum $\kappa$ here is determined by the dominant breakup channel.
The third condition can be shown in a similar manner as Eq.~\eqref{eq:eta-hw}. I apply $H_L$ to this ansatz and find
\begin{equation}
    H_L \mathcal{A}_{L,0}(\vec x) = E_\infty \mathcal{A}_{L,0}(\vec x) +\eta (\vec x)
    \label{eq:diff}
\end{equation}
and
\begin{equation}
   \eta (\vec x) \propto \sum^{\prime}_{\vec n \in {\ZZ}^{d\times A}}     \underbrace{[V(\vec x)-V(\vec x+\vec n L)]}_{\Delta_{\vec n L} V(\vec x)} \mathcal{A}(\vec x+\vec n L)  \,,
    \label{eq:eta}
\end{equation}
where, here and in the following, the primed sum $\sum^{\prime}$ denotes that the term $\vec n = \vec 0$ is omitted. For later convenience, I define the shift and difference operator $\Delta_{\vec n L} $ that shifts an operator $O(\vec x)$ by $\vec x \to \vec x + \vec n L$ and computes the difference with $O(\vec x)$.
With the center of mass fixed at the origin, the localized wave function satisfies $\mathcal{A}(\vec n L) \sim  O[\exp(-\kappa n L)] $ , where $\kappa$ is determined by the dominant breakup channel. Note that for a shift $\vec n$ that translates all particles of a cluster $C$ jointly, the intracluster separations are unchanged, so all interactions internal to the clusters cancel exactly in $\Delta_{\vec n L} V$; only the interaction between the clusters survives, evaluated at separations of order $L$. For the remaining shifts, at least one particle is separated from a bound fragment, and the wave function factor itself supplies the exponential suppression for any bounded potential. Consequently, I only require the interaction between the clusters to be short ranged---a weaker assumption than in Ref.~\cite{Konig:2017krd}, where every interaction is assumed to vanish beyond a strictly finite range. (When the long-range Coulomb force acts between the clusters, it is retained in $H$, and the exponential tails below are understood to be replaced by the corresponding Whittaker tails~\cite{Yu:2022nzm}.) I arrive at
\begin{equation}
    \eta (\vec x) = O\Bigg(\sum^{\prime}_{\vec n \in {\ZZ}^{d\times A}  } \mathcal{A}(\vec x+\vec n L)\Bigg)= O[\exp(- \kappa L)]\,.
\end{equation}
I now take the inner product of Eq.~\eqref{eq:diff} with $\mathcal{A}_{L,0}$ over a single unit cell $B^{d\times A}$ of the torus. (All inner products of periodic wave functions are understood in this sense; over the infinite space they would diverge.) Folding one of the two shifted sums over the unit cells tiles $\RR^{d\times A}$, so $\langle \mathcal{A}_{L,0}|\mathcal{A}_{L,0}\rangle_{B^{d\times A}} = 1 + O[\exp(-\kappa L)]$. I thus obtain the finite volume shift as the following $A\times d$ dimensional overlap integral:
\begin{multline}
 \Delta E_L = \sum^{\prime}_{\vec n \in {\ZZ}^{d\times A}} \int_{B^{d\times A}} \dd \vec x\, \mathcal{A}(\vec x) \Delta_{\vec n L}V(\vec x)  \mathcal{A}(\vec x+\vec n L)\\
 + O\left[\exp(-2\kappa L)\right]\\
 =\sum^{\prime}_{\vec n \in {I}^{d\times A}} \int_{B^{d\times A}} \dd \vec x\, \mathcal{A}(\vec x)  \Delta_{\vec n L} V(\vec x)  \mathcal{A}(\vec x+\vec n L)\\
 + O\left(\exp(-\sqrt{2}\kappa L)\right)\,,
\end{multline}
where $I=\{-1,0,1\}$.
This sum can be arranged into different breakup channels denoted by the tuple $A,C$. On the integration domain $B^{d\times A}$, if I shift $\vec x_{C_1}, \vec x_{C_2},\ldots$, by $L$, then the center of mass for this cluster $C$ is certainly far away from all remaining particles. The localized wave function $\mathcal{A}(\vec x +\vec n L) $ naturally admits a factorization into cluster $C$ and its complement, allowing the label $C$. For a fixed cluster-relative displacement, I group the microscopic shifts that produce the same factorized component under permutations of identical particles. I call the size of this group $G_{A,C}$; geometrically, its members correspond to distinct coordinate-labeled but symmetry-related faces of the periodic domain (cf. the construction illustrated in Figs.~\ref{fig:ite} and \ref{fig:coord}). The above equation can then be rearranged into:
\begin{multline}
\Delta E_L    
 = \sum_{C} G_{A,C} \int_{B^{d\times A}} \dd \vec x\, \mathcal{A}(\vec x)\\
 \times \Delta_{\vec n_C L} V(\vec x) \mathcal{A}(\vec x+\vec n_C L)
 + O\left(\exp(-\sqrt{2}\kappa L)\right)\,.
 \label{eq:overlap}
\end{multline}
With this rearrangement, I switch from the $A$-body discussion to the clustering picture.

The number $G_{A,C}$ is the size of an equivalence class of $L$ shifts. Two shifts are equivalent when they translate the same set of conserved-label species out of the nucleus, so that the resulting factorized states differ only by a permutation of identical particles. Every member of a class contributes the same leading tail: the corresponding overlap integrals are related by such a permutation---whose sign enters squared---together with a relabeling of the integration variables. I may therefore replace each class by a single representative shift $\vec n_C$ and multiply by its size $G_{A,C}$. The factor thus counts the symmetry-related microscopic shifts within the class, as fixed by the conserved labels of the chosen degrees of freedom; it does not count physical breakup channels.
To consider clusters with certain separation involves relative coordinates, and this is where Jacobi lattices become essential. The two-cluster Jacobi coordinates are denoted by $(\vec \rho,\vec \xi, \vec R_0) = (\vec \rho_1, \vec \rho_2,\ldots,$ $ \vec \rho_{A-C-1}, \vec \xi_1 \ldots,\vec \xi_{C-1},\vec R_0)$, with $\rho_i,\xi_i \in \mathbb{R}^d$. Here $\vec \rho_i$ are internal coordinates for the cluster with $A-C$ particles and $\vec \xi_i$ are internal coordinates for the cluster with $C$ particles. The system's COM is fixed at the origin, removing this degree of freedom. $\vec R_0$ is the relative distance between cluster $C$ and cluster $A-C$, and to differentiate from $\rho_i,\xi_i$ in the clusters, I intentionally capitalized it. This transformation to the Jacobi lattice involves a change of integral volume from the Cartesian box $(B\equiv [-L/2,L/2])^{d\times A} $ to the Jacobi box $(B'\equiv [-L/2,L/2])^{d\times (A-1)}$, illustrated in Sec.~\ref{sec:Jacobi_Lattice}.
For $|\vec R_0| \geq L/2$, the wave function factorizes as $\mathcal A(\vec \rho,\vec \xi,\vec R_0) = \psi(\vec R_0) \times \text{remaining}$, with $ \psi(\vec R_0) $ describing the two-body cluster-cluster wave function. In this regime, the breakup is dominated by the leading channel:
\begin{multline}
     \mathcal A(\vec \rho,\vec \xi, \vec R_0) =\mathcal A_{A-C}(\vec \rho_1,\ldots,\vec \rho_{A-C-1}) \\
    \times\mathcal A_C(\vec \xi_1,\ldots,\vec \xi_{C-1})\times \psi(\vec R_0)\\ +O\left[\exp(-\kappa^* L/2)\right]\,,
    \label{eq:clusters}
\end{multline}
where $\kappa^*$ denotes the subleading breakup channels. I assume these $\kappa^*$ are well separated from the leading $\kappa$ for simplicity. The intrinsic fragment states $\mathcal A_{A-C}$ and $\mathcal A_C$ are normalized to unity, and $\psi$ is the cluster-relative factor for one representative microscopic embedding; its asymptotic coefficient defines $C_0$. If several channel components have comparable asymptotic scales, the leading energy shift is a sum over those components, each with its own overlap normalization and geometric factor.
Eq.~\eqref{eq:overlap} in the Jacobi lattice representation now reads:
\begin{multline}
    \Delta E_L = G_{A,C}\int_{(B')^{d\times (A-1)}} \dd \vec \rho\dd \vec \xi\, \dd \vec R_0\, \mathcal{A}(\vec \rho,\vec \xi,\vec R_0) \\ \times\Delta_{\vec n_C L}V(\vec \rho,\vec \xi,\vec R_0)  \mathcal{A}(\vec \rho,\vec \xi,\vec R_0 +\vec n_C L)
    \\ + O\left[\exp(-\kappa^* L)\right]\,,
\end{multline}
where $\vec n_C$ is the dimensionless nearest unit vector ($|\vec n_C| = 1$) of the cluster-relative coordinate $\vec R_0$, so that the physical displacement induced by the original lattice translation is $\vec n_C L$. It serves as the representative of the equivalence class containing $G_{A,C}$ symmetry-related microscopic shifts. My iterative construction guarantees this integral remains unchanged. Since the shift $\vec n_C L$ already specifies the dominant cluster partition, the advantage of the Jacobi lattice transformation is clear: it avoids enumerating all possible cluster coordinate combinations and their associated integral domains.
I will also omit higher exponential terms from now on.
I rewrite the difference potential as $\Delta_{\vec n_C L} V = \Delta_{\vec n_C L} (H - E_\infty)$, where $H(\vec \rho,\vec \xi,\vec R_0 +\vec n_C L)$ has the shifted eigenvector $\mathcal{A}(\vec \rho,\vec \xi,\vec R_0 +\vec n_C L) $, so that the integrand with the $[H(\vec \rho,\vec \xi,\vec R_0 +\vec n_C L) - E_\infty]$ term vanishes [in analogy with Eq.~\eqref{eq:1d-addzero}]:
\begin{multline}
    \Delta E_L = G_{A,C}\int_{(B')^{d\times (A-1)}} \dd \vec \rho\dd \vec \xi\, \dd \vec R_0\, \mathcal{A}(\vec \rho,\vec \xi,\vec R_0) \\
    \times [  H -E_\infty ]   \mathcal{A}(\vec \rho,\vec \xi,\vec R_0 +\vec n_C L) \\
    + O\left[\exp(-\kappa^* L)\right]\,.
\end{multline}
This rewriting is useful because applying Green's second identity,
\begin{equation}
 v \nabla^2 u
 = u \nabla^2 v + \nabla \cdot (v \nabla u - u \nabla v)\,,
\label{eq:simp-int3d}
\end{equation}
converts the overlap integral into a surface term (again, because $\nabla^2$ is not self-adjoint on this domain for these trial functions) plus a volume remainder.
In the region where the cluster factorization Eq.~\eqref{eq:clusters} holds, this gives (for details of this transformation see Ref.~\cite{Yu:2022nzm,Yu:2023ucq}):
\begin{multline}
   \Delta E_L = G_{A,C} \times \\
   \sum_{k = 1}^{(A-1)\times d}\int_{\partial (B')^{d\times (A-1)}} \prod_{i\neq k} \dd s_i  \,
   \frac{1}{2 \mu_k}\bigg[\mathcal A(\vec\rho,\vec\xi,\vec R_0+\vec n_C L)
  \frac{\partial}{\partial s_k} \\ \mathcal A(\vec\rho,\vec\xi,\vec R_0)
  - \mathcal A(\vec\rho,\vec\xi,\vec R_0)
  \frac{\partial}{\partial s_k}\mathcal A(\vec\rho,\vec\xi,\vec R_0+\vec n_C L) \bigg]\\
  + G_{A,C}\int_{(B')^{d\times (A-1)}} \dd \vec \rho\dd \vec \xi\, \dd \vec R_0\, \mathcal{A}(\vec \rho,\vec \xi,\vec R_0 +\vec n_C L) \\
    \times [ H -E_\infty ]   \mathcal{A}(\vec \rho,\vec \xi,\vec R_0 )
\end{multline}
with $s_i$ going through all $\rho_{i,j}$, $\xi_{i,j}$, and $R_{0,j}$ coordinate components in $(\vec \rho, \vec \xi, \vec R_0)$. Here $\partial (B')^{d\times (A-1)}$ denotes the boundary of the full $(A-1)d$-dimensional Jacobi box, and $\mu_k$ are reduced mass factors for these non-normalized Jacobi bases associated with the Laplacian. Most of these terms vanish: first, the volume integral (last two lines) vanishes exactly because $(H- E_\infty)\mathcal{A}(\vec \rho,\vec \xi,\vec R_0) = 0$, i.e., $\mathcal{A}$ is an eigenstate of $H$. Second,
on the cluster-internal boundaries,
the surface terms involving $\partial_{\rho_i}$ and $\partial_{\xi_i}$ vanish because both $\mathcal{A}$ and its shifted copy share the same dependence on internal coordinates, hence $u\nabla v - v \nabla u = 0$ on those faces. Finally, according to Eq.~\eqref{eq:clusters}, the representative Jacobi lattice allows only a shift by $L$ on $\vec R_0$ at the leading exponential order; the $G_{A,C}$ symmetry-related microscopic embeddings have already been collected in the prefactor. The two surviving terms from the Green's identity (involving $\psi \partial \psi' - \psi' \partial \psi$) add constructively on opposite faces of the box, yielding a factor of $2$ that cancels the $1/(2\mu)$ prefactor. This gives:
\begin{multline}
    \Delta E_L =  
\frac{G_{A,C}}{\mu_{A|C}}\sum_{k=1}^d
\int_{\partial (B')^{d\times (A-1)}} \prod_{i\neq k }\dd R_{0,i} 
\prod_{i=1}^{A-C-1} \dd \vec \rho_i 
\\ \prod_{i=1}^{C-1}  \dd \vec \xi_i   |\mathcal A_{A-C}\mathcal A_{C}|^2
    \times   \psi^*(\vec R_0) 
    \frac{\partial}{\partial R_{0,k}}\psi(\vec R_0+\vec n_C L) \\
  =\frac{ G_{A,C}}{\mu_{A|C}}\sum_{k=1}^d\int_{\partial (B')^{d}} \prod_{i\neq k }\dd R_{0,i}\\
    \times \psi^*(\vec R_0)
    \frac{\partial}{\partial R_{0,k}}\psi(\vec R_0+\vec n_C L)\,,
    \label{eq:surface-cluster}
\end{multline}
where $R_{0,k}$ is the $k$th component of $\vec R_0$, and $\mu_{A|C} = C(A-C)m/A$ is the reduced mass of the cluster pair, associated with the Laplacian in $\vec R_0$.
This is the surface form of Ref.~\cite{Yu:2022nzm}.  For   $d = 3$ and $l=0$, substituting the asymptotic Whittaker form I obtain
\begin{multline}
    \Delta E_L  \approx G_{A,C}  \times \left[{-}\frac{3\anc^2}{\mu_{A|C} L}
  \Big[W_{-\bar\eta,\frac{1}{2}}(\kappa L)\Big]^2\right]\\
  = G_{A,C}\, \Delta E_{\textrm{TB},L}\,,
\end{multline}
i.e., the two-body result Eq.~\eqref{eq:DeltaE-3D-final} evaluated with the cluster parameters $\mu_{A|C}$, $\kappa$, and $\anc$.
The geometric factor thus arises from the multiplicity of symmetry-related microscopic embeddings of a fixed shift $\vec n_C L$ on the hypertorus, a direct consequence of describing the system in nucleon degrees of freedom. It is distinct from the factor $3=d$ already present in the two-body result, which counts the spatial directions of the leading cluster-relative faces in Eq.~\eqref{eq:surface-cluster}. Lüscher's formula, when generalized to cluster configurations, naturally yields $G_{A,C}$, provided one carefully accounts for permutations when selecting the cluster indices. The physical system is of course unaware of this labeling, and the lattice Hamiltonian does not involve this choice until clusters are formed. My Jacobi lattice construction directly connects the Cartesian and Jacobi descriptions in Eq.~\eqref{eq:overlap} and makes the geometric factor explicit.
Counting this multiplicity factor is straightforward: it runs over the conserved single-nucleon labels. When the interaction conserves the individual spin projections, a cluster with composition $C = (C_{p\uparrow},C_{p\downarrow},C_{n\uparrow},C_{n\downarrow})$ in a  nucleus $A = (A_{p\uparrow},A_{p\downarrow},A_{n\uparrow},A_{n\downarrow})$ has
\begin{equation}
    G_{A,C} = \binom{A_{p\uparrow}}{C_{p\uparrow}}\binom{A_{p\downarrow}}{C_{p\downarrow}}\binom{A_{n\uparrow}}{C_{n\uparrow}}\binom{A_{n\downarrow}}{C_{n\downarrow}}\,,
\label{eq:G-spin}
\end{equation}
which gives the number of faces that contribute at the leading order.
The SU(4)-invariant interaction used in Sec.~\ref{sec:numerics} conserves these labels: 
it contains no tensor or spin-orbit components, so the spin-isospin indices are spectators on which
the strong interaction acts as the identity.
Tensor and spin-orbit forces end this spectator role for the spin indices: 
they mix the spin projections, which are then no longer good labels for this 
count. The isospin indices (for the interactions with pions integrated out) remain spectators 
used to construct the basis, 
consistent with charge conservation. The face counting is resolved only by proton
and neutron species:
\begin{equation}
    G_{A,C} = \binom{A_{p}}{C_{p}}\binom{A_{n}}{C_{n}}\,.
\label{eq:G-nospin}
\end{equation}
The mechanism behind this change is geometric: each face of the 
many-body torus decays with the breakup momentum of the lowest 
configuration \textit{available on that face}. When the individual 
spin projections are conserved, only faces whose composition matches 
the fragment ground states reach the smallest $\kappa$---separating
a four-nucleon set with two spin-up protons from \isotope[16]{O}, for 
example, does not project onto the $\alpha$ ground state and is 
exponentially suppressed. Tensor and spin-orbit forces make the 
spin projections dynamical; the fragment ground states then 
contain all spin compositions compatible with the conserved 
totals, and every selection of $C_p$ protons and $C_n$ neutrons 
carries the same leading tail. The physical breakup channels are 
unchanged in number and identity throughout---what changes is which
faces of the configuration space reach the lowest threshold, a 
distinction no channel-counting rule can capture. 
A similar analysis applies whenever finer degrees of freedom are 
resolved, e.g., quarks in lattice QCD, where a different factor should
appear.

I emphasize that $G_{A,C}$ is not a multiplicity of physical channels. Permuting identical nucleons within a fixed cluster component does not produce additional breakup channels, because the system is unaware of which labeled nucleons form the cluster. A physical-channel count therefore cannot produce this factor. Nor can a channel argument explain why particle statistics leave the factor unchanged. In the overlap integral Eq.~\eqref{eq:overlap}, the $G_{A,C}$ shifts related by permutations of identical particles contribute equally, and any permutation sign enters squared, so symmetric and antisymmetric wave functions yield the same counting. The combinatorial factor of Ref.~\cite{Konig:2017krd} was asserted for identical particles without derivation. The genuine many-body overlap integral makes the counting rule explicit and shows that it is set by the conserved quantum labels of the chosen degrees of freedom rather than by the physical channels. In this sense the factor is a geometric property of the many-body configuration space.
This factor can also be large: for a symmetric split $C = A/2$, it grows combinatorially with $A$, though this growth can sometimes be compensated by large breakup energies. For \isotope[100]{Sn}, where NLEFT calculations have recently been extended to~\cite{Niu:2025uxk}, I expect $G_{\isotope[100]{Sn},\alpha} = 25^4 \approx \ee ^{13}$, using the spin-resolved counting of Eq.~\eqref{eq:G-spin} appropriate for the SU(4)-type interactions employed there. This is no small number especially when the volume correction scales as $\Delta E_L \sim \exp(-\kappa L)$ with typical $L \sim 10$, $\kappa \sim 0.7$ in lattice units, owing to computational limitations.  As shown below, this large factor can sometimes overtake the nominally leading channel in the volume dependence, making it difficult to interpret asymptotic parameters. Moreover, one can expect an even larger $ G_{A,C}$ for simulations with mutable spins or isospins.
\bigskip
\section{Implications for the scattering region}
\label{sec:scattering}

Given that the geometric factor can grow rapidly with the number of degrees of freedom, one may worry that it similarly affects the extraction of scattering observables.
Indeed, the volume dependence of energy levels in the scattering region could suffer from the same amplification owing to the expanding configuration space. However, the observable of interest in this region---the phase shift---is given by the quantization condition derived from matching coefficients in different bases (Jost solutions vs truncated bases). Since the geometric factor multiplies both the numerator and denominator of $\cot[\delta_l (p)] = a_l/b_l$ in Ref.~\cite{Luscher:1990ux}, it cancels in the ratio. I therefore expect existing scattering formulations to remain valid. However, how geometric factors impact scattering matrices near poles deserves further attention, to validate whether a similar factor could reappear.

\bigskip

\section{Numerical verifications}
\label{sec:numerics}
\begin{figure}
    \centering
    \includegraphics[width=1\linewidth]{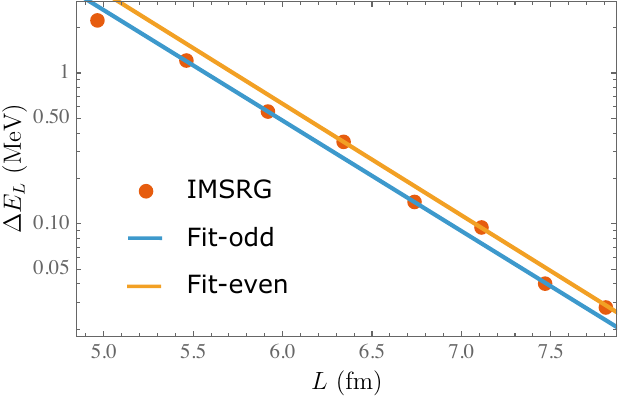}
    \caption{Truncation dependence in IMSRG calculations. The simple \isotope[4]{He} system serves as a validation case. Truncations of odd and even $e_{\rm max} $ are separated to extract ANCs. I ignored the contribution of the Coulomb force, and extracted ANCs for  the $\isotope[4]{He}-N$ cluster, where $N$ is any nucleon, using the simple exponential form of truncation dependence. }
    \label{fig:he-imsrg}
\end{figure}

I provide numerical evidence for my derivation of $G_{A,C}$ in this section. Throughout this section, $\ln$ denotes the natural logarithm (base $e$).
To verify my geometric factor in the single-particle HO basis, I perform IMSRG calculations on \isotope[4]{He} using the $\Delta$-full interaction with cutoff $\Lambda = 394\, \rm MeV$~\cite{Ekstrom:2017koy,Miyagi:2023qce}. I use $L \approx \sqrt{2 e_{\rm max} +7 } b$ from Ref.~\cite{Furnstahl:2013vda}, with $b$ the oscillator length. There have been numerous calculations on ANCs for \isotope[4]{He}, e.g., Ref.~\cite{Nollett:2011qf}, and it is a good benchmark to see if there is an $A$ factor in the single-particle frame.
Because $e_{\rm max}$ is defined by $2 n + l$, odd $e_{\rm max}$ truncates the system differently compared to even $e_{\rm max}$. For this IR extrapolation of \isotope[4]{He}, I separate them in the numerical fitting.
I find in Fig.~\ref{fig:he-imsrg} that $C_{\isotope[4]{He},N} = 6.9(3) \, \rm fm^{-1/2}$ from fitting even harmonic quanta $e_{\rm max}= 6,8,10$, and 
$C_{\isotope[4]{He},N} = 5.8(1) \, \rm fm^{-1/2}$
from fitting odd $e_{\rm max}= 5,7,9$, and I use this as my range of ANCs from the simple exponential formula~\cite{More:2013rma}. Neglecting the Coulomb interaction is well justified for \isotope[4]{He}: the Sommerfeld parameter of the $\isotope[3]{H}$--$p$ channel is only $\bar\eta \approx 0.03$. Without the $A$ factor, the same fit would require IR corrections smaller by a factor of $A$ to retrieve $C_{\isotope[4]{He},N} \approx 6\, \rm fm^{-1/2}$. Because this benchmark is technically simple and IMSRG truncation effects are not expected to change the correction by a factor of $A$, this test shows that the geometric factor is already present in HO-basis extrapolations used with IMSRG and coupled-cluster calculations. It also suggests that ANCs can be extracted accurately once the IR formula includes this factor.

\begin{figure}
    \centering
    \includegraphics[width=1\linewidth]{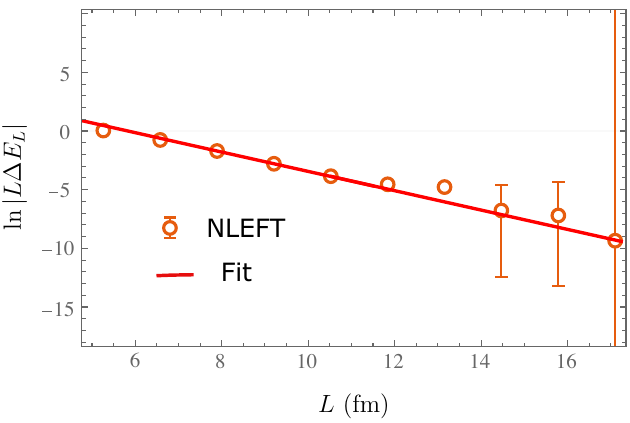}
    \caption{Volume dependence {in} NLEFT calculations on \isotope[4]{He}. In this calculation the SU(4) invariant essential interaction is used~\cite{Lu:2018bat}. I ignored the contribution of the Coulomb force, and extracted ANCs for  $\isotope[4]{He}-N$ cluster, where $N$ is any nucleon, using the simple exponential form of volume dependence. The logarithmic ordinate uses $L|\Delta E_L|$.}
    \label{fig:henleft}
\end{figure}

\begin{figure}
    \centering
    \includegraphics[width=1\linewidth]{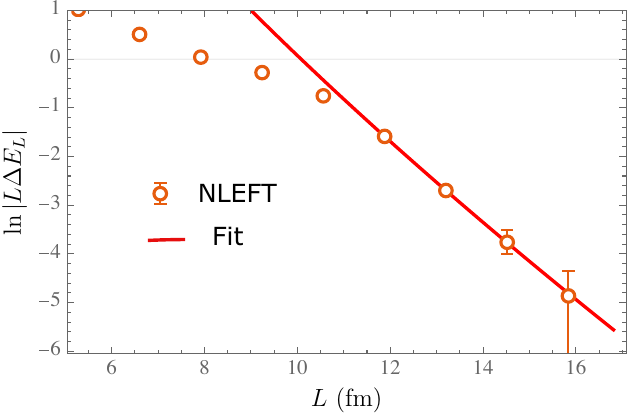}
     \caption{Volume dependence in NLEFT calculations on \isotope[20]{Ne}. In this calculation, the same SU(4) invariant essential interaction is used as in Fig.~\ref{fig:henleft}. Coulomb interaction between clusters is nonperturbative ($\bar\eta >1$) in this case, so I use Eq.~\eqref{eq:DeltaE-3D-final}.  I extract ANCs for  $\isotope[20]{Ne}-\alpha$ cluster.}
    \label{fig:ne20}
\end{figure}

For my main result in Sec.~\ref{sec:main}, I use NLEFT calculations with varied lattice volume to extract ANCs as my benchmark for periodic boundaries. Reviews of this method can be found in Refs.~\cite{Lee:2008fa,Lahde:2019npb}. The interaction I use here is an essential SU(4) invariant interaction~\cite{Lu:2018bat}, which can accurately produce ground states for most light and medium-mass alpha-clustering nuclei. The first two choices are \isotope[4]{He} (Fig.~\ref{fig:henleft}) and \isotope[20]{Ne} (Fig.~\ref{fig:ne20}). This choice is based on the available ANC benchmarks for \isotope[4]{He}, and in a recent work  \isotope[20]{Ne} ANC is found to be very stable across different methods~\cite{Harris:2025zsh}.  Lattice calculations presented here use the $O(a^3)$ improved kinetic operator with lattice spacing $a = 1.32$ fm~\cite{Lu:2018bat}. Owing to the simple interactions, \isotope[20]{Ne} breakup energy is around 1.5 MeV, making the \isotope[16]{O}-$\alpha$
channel the one with the smallest breakup momenta compared to the actual proton channel. The experimental breakup energy of  \isotope[20]{Ne} is at 4.7 MeV. Details of the numerical fitting can be found in Ref.~\cite{Yu:2022nzm}. Throughout this section, the parenthetical uncertainties---including those in Table~\ref{tab:ancs}---are confidence intervals obtained from weighted fits that incorporate the Monte Carlo uncertainty of each energy level, propagated to the ANC by resampling. These resampled confidence intervals provide a direct measure of the statistical fit uncertainty. In each fit, both the ANC $C_0$ and the binding momentum $\kappa$ are free parameters, using the simple exponential form for \isotope[4]{He} and the Whittaker-squared form of Eq.~\eqref{eq:DeltaE-3D-final} for the Coulombic channels. Only the subleading \isotope[15]{N}--$p$ channel momentum in the \isotope[16]{O} fit is fixed from its threshold. I find that
$C_{\isotope[4]{He},N} = 6.7(2) \, \rm fm^{-1/2}$, 
$C_{\isotope[20]{Ne},\alpha} = 4.2(2)\times 10^3 \, \rm fm^{-1/2}$.
Without this factor, the same fits would yield
$C_{\isotope[4]{He},N} = 13.4(4) \, \rm fm^{-1/2}$, 
$C_{\isotope[20]{Ne},\alpha} = 1.05(5)\times 10^5 \, \rm fm^{-1/2}$, which are clearly incompatible with the benchmark scale. The pinhole method value in Ref.~\cite{Harris:2025zsh} is fine tuned so that breakup energy is closer to the physical value. I also expect fine-tuned calculations to move even closer to the pinhole results.

\begin{figure}
    \centering
    \includegraphics[width=1\linewidth]{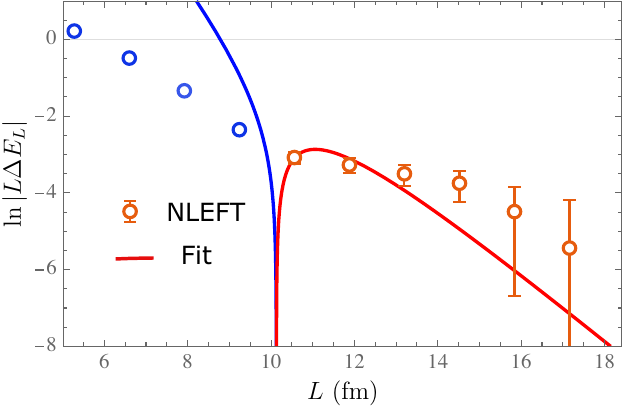}
    \caption{Geometric factors in NLEFT calculations on \isotope[16]{O}. In this calculation, the same SU(4) invariant essential interaction is used as in Fig.~\ref{fig:henleft}. In addition to the nonperturbative Coulomb interaction between clusters, the leading channel at very large $L$ is in fact $\isotope[15]{N} - p$. However, the geometric factor  $G_{\isotope[16]{O},\alpha} = 256 $ gives a significant boost to the $\isotope[16]{O}-\alpha$ channel and {creates} a change of sign in {the} volume dependence at $L \approx 10~\rm fm$. The logarithmic ordinate uses $L|\Delta E_L|$, with blue indicating negative shifts and red indicating positive shifts. The fit range is $10$--$16~\rm fm$.}
    \label{fig:16O}
\end{figure}

\begin{figure}[t!]
    \centering
    \includegraphics[width=1\linewidth]{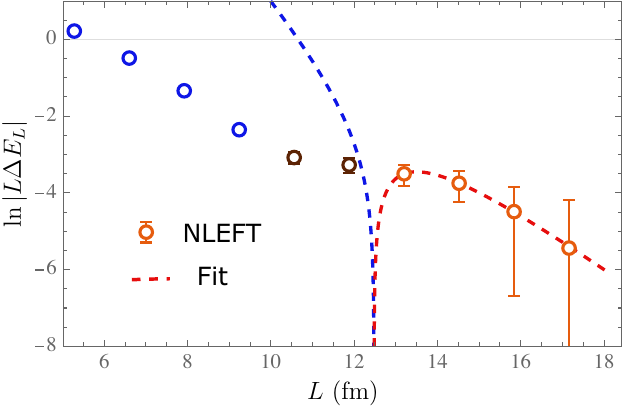}
    \caption{Geometric factors in NLEFT calculations on \isotope[16]{O}. All parameters are the same as in Fig.~\ref{fig:16O}, except that only the last four data points ($L \approx 13$--$17~\rm fm$) enter the fit. This figure illustrates the sensitivity to the Monte Carlo noise.}
    \label{fig:16O-range}
\end{figure}

I then proceed to check an interesting case: \isotope[16]{O}. For \isotope[16]{O} , the $\alpha$ breakup channel is one of the subleading channels compared to the proton breakup channel when breakup momenta $\kappa$ are concerned. I find  $\kappa _p \approx 150 \, \rm MeV$ and  $\kappa _\alpha \approx 190\, \rm MeV$ from experimental data~\cite{Harris:2025zsh}. In my NLEFT calculation I also find that $B_\alpha = 6.7\, \rm {MeV}$, very close to the experimental $B_\alpha = 7.2\,\rm MeV$. In the volume dependence, negative corrections to the energy dominate in the conventionally accessible volumes $L \lesssim 10$ fm, indicating an even-parity leading contribution. However, for \isotope[16]{O}, the leading proton channel should produce a positive correction according to Ref.~\cite{Konig:2011nz} because \isotope[15]{N}'s ground state has $J^\pi = {1/2^-}$. This makes \isotope[16]{O} a useful stress test of the counting factor: $G_{\isotope[16]{O},\alpha} = 256 $ can amplify the subleading $\alpha$ contribution relative to the proton contribution with $G_{\isotope[16]{O},p} = 8 $. Using a two-channel approximation (on the numerically subleading proton channel),
\begin{multline}
    \Delta E_L  \approx {-}\frac{3 G_{\isotope[16]{O},\alpha} C_{\isotope[16]{O},\alpha}^2}{\mu_{\isotope[16]{O}|\alpha} L}\Big[W_{-\bar\eta_{\alpha},\frac{1}{2}}(\kappa_{\alpha} L)\Big]^2\\
    +\frac{3G_{\isotope[16]{O},p} C_{\isotope[16]{O},p}^2}{\mu_{\isotope[16]{O}|p} L}
  \Big[W_{-\bar\eta_p,\frac{3}{2}}(\kappa_{p} L)\Big]^2\,.
\end{multline}
I estimate in Fig.~\ref{fig:16O} $C_{\isotope[16]{O},\alpha} = 350(50) \, \rm fm^{-1/2}$, consistent within uncertainties with the pinhole estimation $C_{\isotope[16]{O},\alpha} = 380(80)\, \rm fm^{-1/2}$~\cite{Harris:2025zsh}. Without this amplification, the present volume dependence would not support a stable extraction of the $\alpha$-channel ANC for \isotope[16]{O}, and even the qualitative volume dependence would be different. However, in this case I notice extreme sensitivity to the range of data selected to extract ANCs.  For example, in Fig.~\ref{fig:16O-range}, fitting the last four available data points yields $C_{\isotope[16]{O},\alpha} = 900(200)\, \rm fm^{-1/2}$, likely  because the Monte Carlo statistics need to be improved before consistent results can be obtained for \isotope[16]{O}. Moreover, \isotope[15]{N} is an odd-even {nucleus} and is not accurately {described} by the essential interaction used in this paper. I therefore regard the \isotope[16]{O} analysis primarily as a demonstration that the geometric factor changes the sign of the volume dependence in the accessible volumes---a qualitative feature that no choice of fit window can undo---rather than as a precision extraction of the ANC. I expect the new wave function matching method~\cite{Elhatisari:2022zrb} to help identify this inconsistency and provide more consistent results for understanding the \isotope[12]{C}$(\alpha,\gamma)$\isotope[16]{O} reaction. Table~\ref{tab:ancs} summarizes the ANCs extracted in this work.

\begin{table*}[t]
\caption{Summary of ANCs extracted in this work. The column ``without $G_{A,C}$'' repeats the same fits with the geometric factor set to one, which rescales the ANC by $\sqrt{G_{A,C}}$; for the $\isotope[16]{O}$--$\alpha$ rows, the $\alpha$ channel then becomes subleading and no stable ANC can be extracted (see text). For the $\isotope[4]{He}$--$N$ breakup, the four one-nucleon channels are (assumed) degenerate, and $G_{A,C}$ denotes their combined multiplicity. All ANCs are in units of $\rm fm^{-1/2}$.}
\label{tab:ancs}
\begin{ruledtabular}
\begin{tabular}{llcccc}
System & Method & $G_{A,C}$ & ANC with $G_{A,C}$ & ANC without $G_{A,C}$ & Reference \\
\hline
$\isotope[4]{He}$--$N$ & IMSRG (HO basis), even $e_{\rm max}$ & $4$ & $6.9(3)$ & $13.8(6)$ & $\approx 6$~\cite{Nollett:2011qf} \\
$\isotope[4]{He}$--$N$ & IMSRG (HO basis), odd $e_{\rm max}$ & $4$ & $5.8(1)$ & $11.6(2)$ & $\approx 6$~\cite{Nollett:2011qf} \\
$\isotope[4]{He}$--$N$ & NLEFT, $L \approx 7\text{--}16$ fm& $4$ & $6.7(2)$ & $13.4(4)$ & $\approx 6$~\cite{Nollett:2011qf} \\
$\isotope[20]{Ne}$--$\alpha$ & NLEFT, $L\approx 11\text{--}16$ fm & $625$ & $4.2(2)\times 10^3$ & $1.05(5)\times 10^5$ & 3800(950)~\cite{Harris:2025zsh} \\
$\isotope[16]{O}$--$\alpha$ & NLEFT, $L \approx 10\text{--}16$ fm & $256$ & $350(50)$ & subleading & $380(80)$~\cite{Harris:2025zsh} \\
$\isotope[16]{O}$--$\alpha$ & NLEFT, $L \approx 13 \text{--}17$ fm & $256$ & $900(200)$ & subleading & $380(80)$~\cite{Harris:2025zsh} \\
\end{tabular}
\end{ruledtabular}
\end{table*}

\section{Conclusions}
In summary, I have derived the geometric factor $G_{A,C}$ that has been missing when volume dependence formulas are applied to study finite nuclei. Accounting for this factor is necessary for extracting correct ANCs in finite-volume methods inspired by Lüscher. It also implies that microscopic calculations of nuclear reactions can have larger finite-volume effects than expected from point-cluster formulas. As seen in \isotope[20]{Ne}, breakup contributions in finite volume can be strongly amplified, making it difficult to fit asymptotic wave function tails in the relatively small volumes accessible to numerical simulations. The Coulomb tail enhances this dependence further. Nonperturbative Coulomb corrections, as shown for pointlike clusters in Ref.~\cite{Yu:2022nzm}, take a Whittaker-squared form. At low energies or for strong Coulomb interactions ($\kappa L \lesssim 4\bar\eta$), this form decays as $\exp(-4\sqrt{\gamma L})$, with $\gamma = \kappa \bar\eta = \mu Z_1 Z_2$ the inverse Bohr radius of the cluster pair: the decay is then set by the Coulomb barrier alone and becomes independent of the binding momentum. Breakup channels with larger $\kappa^*$ but comparable charge products therefore lose their relative exponential suppression, and calculations that ignore amplified subleading breakup channels may misidentify the dominant source of the finite-volume energy shift.

Furthermore,  ANCs are very sensitive to the breakup threshold. Experimental groups aim to reduce errors to the percent level~\cite{deBoer:2017ldl} to fully understand nucleosynthesis in stellar environments. A common practical solution to this sensitivity is parameter tuning to match the infinite-volume energy. However, this tuning partly weakens the predictive character of first-principles calculations. Independently of direct ANC extraction, the geometric factor grows for heavier clusters and can strongly amplify breakup energy shifts. I conclude that this factor needs to be carefully accounted for as computational power increases, to predict both the energy structures and the ANCs more reliably from an \textit{ab initio} perspective. A recently published result supports this point. Reference~\cite{Niu:2025uxk} provides Coulomb-free volume correction data for \isotope[40]{Ca} and \isotope[100]{Sn}. From my result, I estimate that $G_{ \isotope[100]{Sn},\alpha } \approx 40 \times G_{ \isotope[40]{Ca},\alpha }$, so this factor alone should produce a significant amplification. Indeed, one observes an increase by a factor of approximately 30. This signals that $G_{A,C}$ alone, which amplifies the alpha-cluster ANCs, plays an important role for lattice calculations of heavier elements. I expect that the same effect will appear in any many-body methods that have emergent cluster degrees of freedom.

Finally, while the SU(4)-invariant interaction used here conserves the individual spin projections, realistic nuclear forces contain tensor and spin-orbit components that mix them. With the normalization fixed by Eq.~\eqref{eq:clusters}, my derivation predicts that the counting coarsens from Eq.~\eqref{eq:G-spin} to Eq.~\eqref{eq:G-nospin}. For the $\isotope[16]{O}$--$\alpha$ channel, $G$ increases from $256$ to $784$. This would shift the extracted ANC by a factor of $\sqrt{256/784} \approx 0.57$. By construction, this difference cannot be resolved within the spin-conserving simulations presented here. Testing the prediction requires interactions with explicit tensor and spin-orbit components, which I leave to future work with high-fidelity interactions.

\begin{acknowledgments}
I thank Sebastian König and Dean Lee for enlightening discussions. I thank Yuanzhuo Ma for providing part of the lattice data for numerical analysis. This work is in part supported by JST ERATO Grant No. JPMJER2304, Japan. This work is also in part supported by the Multidisciplinary Cooperative Research Program in CCS, University of Tsukuba.
\end{acknowledgments}
\section*{Data Availability}
There are no publicly available research data or software supporting this paper. Requests for further information or data should be sent to the author.
\bibliography{refs.bib}
\end{document}

%% file: defs.tex
\newcommand{\apriori}{\textit{a priori}\xspace}
\WithSuffix\newcommand\apriori*{\textit{a-priori}\xspace}

\newcommand{\ee}{\mathrm{e}}

\newcommand{\ZZ}{\mathbb{Z}}
\newcommand{\RR}{\mathbb{R}}

\newcommand*\rvec[1]%
{\ensuremath{\overset{\smash{\raisebox{-1.5pt}{\tiny$\rightarrow$}}}{#1}}}
\newcommand*\lvec[1]%
{\ensuremath{\overset{\smash{\raisebox{-1.5pt}{\tiny$\leftarrow$}}}{#1}}}

\NewEnviron{subalign}[1][]{%
\begin{subequations}\begin{align}
  \BODY
\end{align}\label{#1}\end{subequations}
}

\NewEnviron{spliteq}{%
\begin{equation}\begin{split}
  \BODY
\end{split}\end{equation}
}

\NewEnviron{mleq}{%
\begin{equation}
  \BODY
\end{equation}
}

\renewcommand{\vec}[1]{\mathbf{#1}}

\newcommand{\anc}{C_0}
\renewcommand{\vec}[1]{\mathbf{#1}}
\renewcommand{\binom}[2]{\genfrac{(}{)}{0pt}{}{#1}{#2}}

%% file: refs.bib
@article{Miyagi:2023qce,
    author = "Miyagi, Takayuki",
    title = "{NuHamil : A numerical code to generate nuclear two- and three-body matrix elements from chiral effective field theory}",
    eprint = "2302.07962",
    archivePrefix = "arXiv",
    primaryClass = "nucl-th",
    doi = "10.1140/epja/s10050-023-01039-y",
    journal = "Eur. Phys. J. A",
    volume = "59",
    number = "7",
    pages = "150",
    year = "2023"
}

@article{Hergert:2020bxy,
    author = "Hergert, H.",
    title = "{A Guided Tour of $ab$ $initio$ Nuclear Many-Body Theory}",
    eprint = "2008.05061",
    archivePrefix = "arXiv",
    primaryClass = "nucl-th",
    reportNumber = "INT-PUB-20-032",
    doi = "10.3389/fphy.2020.00379",
    journal = "Front. in Phys.",
    volume = "8",
    pages = "379",
    year = "2020"
}

@article{Konig:2017krd,
    author = {K\"onig, Sebastian and Lee, Dean},
    title = "{Volume Dependence of N-Body Bound States}",
    eprint = "1701.00279",
    archivePrefix = "arXiv",
    primaryClass = "hep-lat",
    doi = "10.1016/j.physletb.2018.01.060",
    journal = "Phys. Lett. B",
    volume = "779",
    pages = "9--15",
    year = "2018"
}

@book{Lahde:2019npb,
    author = {L\"ahde, Timo A. and Mei\ss{}ner, Ulf-G.},
    title = "{Nuclear Lattice Effective Field Theory}: {An introduction}",
    doi = "10.1007/978-3-030-14189-9",
    isbn = "978-3-030-14187-5, 978-3-030-14189-9",
    publisher = "Springer",
    volume = "957",
    year = "2019"
}

@article{Luscher:1990ux,
    author = "Luscher, Martin",
    title = "{Two particle states on a torus and their relation to the scattering matrix}",
    reportNumber = "DESY-90-131",
    doi = "10.1016/0550-3213(91)90366-6",
    journal = "Nucl. Phys. B",
    volume = "354",
    pages = "531--578",
    year = "1991"
}

@article{Ekstrom:2015rta,
    author = {Ekstr\"om, A. and Jansen, G. R. and Wendt, K. A. and Hagen, G. and Papenbrock, T. and Carlsson, B. D. and Forss\'en, C. and Hjorth-Jensen, M. and Navr\'atil, P. and Nazarewicz, W.},
    title = "{Accurate nuclear radii and binding energies from a chiral interaction}",
    eprint = "1502.04682",
    archivePrefix = "arXiv",
    primaryClass = "nucl-th",
    doi = "10.1103/PhysRevC.109.059901",
    journal = "Phys. Rev. C",
    volume = "91",
    number = "5",
    pages = "051301",
    year = "2015",
    note = "[Erratum: Phys.Rev.C 109, 059901 (2024)]"
}

@article{Machleidt2011,
    author = "Machleidt, R. and Entem, D. R.",
    title = "{Chiral effective field theory and nuclear forces}",
    eprint = "1105.2919",
    archivePrefix = "arXiv",
    primaryClass = "nucl-th",
    doi = "10.1016/j.physrep.2011.02.001",
    journal = "Phys. Rept.",
    volume = "503",
    pages = "1--75",
    year = "2011"
}

@article{Elhatisari:2022zrb,
    author = "Elhatisari, Serdar and others",
    title = "{Wavefunction matching for solving quantum many-body problems}",
    eprint = "2210.17488",
    archivePrefix = "arXiv",
    primaryClass = "nucl-th",
    doi = "10.1038/s41586-024-07422-z",
    journal = "Nature",
    volume = "630",
    number = "8015",
    pages = "59--63",
    year = "2024"
}

@article{Ekstrom:2017koy,
    author = {Ekstr\"om, A. and Hagen, G. and Morris, T. D. and Papenbrock, T. and Schwartz, P. D.},
    title = "{$\Delta$ isobars and nuclear saturation}",
    eprint = "1707.09028",
    archivePrefix = "arXiv",
    primaryClass = "nucl-th",
    doi = "10.1103/PhysRevC.97.024332",
    journal = "Phys. Rev. C",
    volume = "97",
    number = "2",
    pages = "024332",
    year = "2018"
}

@article{Hagen:2013nca,
    author = "Hagen, G. and Papenbrock, T. and Hjorth-Jensen, M. and Dean, D. J.",
    title = "{Coupled-cluster computations of atomic nuclei}",
    eprint = "1312.7872",
    archivePrefix = "arXiv",
    primaryClass = "nucl-th",
    doi = "10.1088/0034-4885/77/9/096302",
    journal = "Rept. Prog. Phys.",
    volume = "77",
    number = "9",
    pages = "096302",
    year = "2014"
}

@article{Konig:2011nz,
  title = {Volume {Dependence} of {Bound} {States} with {Angular} {Momentum}},
  volume = {107},
  doi = {10.1103/PhysRevLett.107.112001},
  journal = {Phys. Rev. Lett.},
  author = {König, Sebastian and Lee, Dean and Hammer, H.-W.},
  year = {2011},
  pages = {112001},
}

@article{Luscher:1985dn,
  title = {Volume dependence of the energy spectrum in massive quantum field
theories - {I}. {Stable} particle states},
  volume = {104},
  issn = {1432-0916},
  doi = {10.1007/BF01211589},
  number = {2},
  journal = {Comm. Math. Phys.},
  author = {Lüscher, M.},
  year = {1986},
  pages = {177--206},
}

@article{Luscher:1986pf,
  title = {Volume dependence of the energy spectrum in massive quantum field
theories - {II}. {Scattering} states},
  volume = {105},
  issn = {1432-0916},
  doi = {10.1007/BF01211097},
  number = {2},
  journal = {Comm. Math. Phys.},
  author = {Lüscher, M.},
  year = {1986},
  pages = {153--188},
}

@article{Konig:2020lzo,
  title = {Few-{{Body Bound States}} and {{Resonances}} in {{Finite Volume}}},
  author = {K{\"o}nig, Sebastian},
  year = {2020},
  journal = {Few-Body Syst.},
  volume = {61},
  number = {3},
  pages = {20},
  issn = {1432-5411},
  doi = {10.1007/s00601-020-01550-8}
}

@article{Luscher:1991cf,
  title = {Signatures of Unstable Particles in Finite Volume},
  author = {L{\"u}scher, Martin},
  year = {1991},
  journal = {Nuclear Physics B},
  volume = {364},
  number = {1},
  pages = {237--251},
  issn = {0550-3213},
  doi = {10.1016/0550-3213(91)90584-K}
}

@article{More:2013rma,
  title = {Universal Properties of Infrared Oscillator Basis Extrapolations},
  author = {More, S. N. and Ekstr{\"o}m, A. and Furnstahl, R. J. and Hagen, G.
and Papenbrock, T.},
  year = {2013},
  journal = {Phys. Rev. C},
  volume = {87},
  number = {4},
  pages = {044326},
  doi = {10.1103/PhysRevC.87.044326}
}

@article{Furnstahl:2013vda,
  title = {Systematic Expansion for Infrared Oscillator Basis Extrapolations},
  author = {Furnstahl, R. J. and More, S. N. and Papenbrock, T.},
  year = {2014},
  journal = {Phys. Rev. C},
  volume = {89},
  number = {4},
  pages = {044301},
  doi = {10.1103/PhysRevC.89.044301}
}

@article{Furnstahl:2014hca,
  title = {Infrared Extrapolations for Atomic Nuclei},
  author = {Furnstahl, R. J. and Hagen, G. and Papenbrock, T. and Wendt, K. A.},
  year = {2015},
  journal = {J. Phys. G: Nucl. Part. Phys.},
  volume = {42},
  number = {3},
  pages = {034032},
  issn = {0954-3899},
  doi = {10.1088/0954-3899/42/3/034032}
}

@article{Briceno:2012rv,
    author = "Briceno, Raul A. and Davoudi, Zohreh",
    title = "{Three-particle scattering amplitudes from a finite volume
formalism}",
    eprint = "1212.3398",
    archivePrefix = "arXiv",
    primaryClass = "hep-lat",
    doi = "10.1103/PhysRevD.87.094507",
    journal = "Phys. Rev. D",
    volume = "87",
    number = "9",
    pages = "094507",
    year = "2013"
}

@article{Polejaeva:2012ut,
    author = "Polejaeva, K. and Rusetsky, A.",
    title = "{Three particles in a finite volume}",
    eprint = "1203.1241",
    archivePrefix = "arXiv",
    primaryClass = "hep-lat",
    doi = "10.1140/epja/i2012-12067-8",
    journal = "Eur. Phys. J. A",
    volume = "48",
    pages = "67",
    year = "2012"
}

@article{Hammer:2017kms,
  author = {Hammer, H.-W. and Pang, Jin-Yi and Rusetsky, Akaki},
  title = {Three-particle quantization condition in a finite volume: 2. General
formalism and the analysis of data},
  journal = {Journal of High Energy Physics},
  year = {2017},
  month = {Oct},
  day = {17},
  volume = {2017},
  number = {10},
  pages = {115},
  issn = {1029-8479},
  doi = {10.1007/JHEP10(2017)115}
}

@article{Meissner:2014dea,
  author         = "Meißner, Ulf-G. and Ríos, Guillermo and Rusetsky,
                    Akaki",
  title          = "{Spectrum of three-body bound states in a finite volume}",
  journal        = "Phys. Rev. Lett.",
  volume         = "114",
  year           = "2015",
  number         = "9",
  pages          = "091602",
  doi            = "10.1103/PhysRevLett.117.069902,
                    10.1103/PhysRevLett.114.091602",
  note           = "[Erratum: Phys. Rev. Lett.117 069902 (2016)]",
  eprint         = "1412.4969",
  archivePrefix  = "arXiv",
  primaryClass   = "hep-lat"
}

@article{Briceno:2019muc,
  author         = "Briceño, Raúl A. and Hansen, Maxwell T. and Sharpe,
                    Stephen R. and Szczepaniak, Adam P.",
  title          = "{Unitarity of the infinite-volume three-particle
                    scattering amplitude arising from a finite-volume
                    formalism}",
  journal        = "Phys. Rev. D",
  volume         = "100",
  year           = "2019",
  number         = "5",
  pages          = "054508",
  doi            = "10.1103/PhysRevD.100.054508",
  eprint         = "1905.11188",
  archivePrefix  = "arXiv",
  primaryClass   = "hep-lat"
}

@article{Mai:2017bge,
  author         = "Mai, M. and Döring, M.",
  title          = "{Three-body Unitarity in the Finite Volume}",
  journal        = "Eur. Phys. J. A",
  volume         = "53",
  year           = "2017",
  number         = "12",
  pages          = "240",
  doi            = "10.1140/epja/i2017-12440-1",
  eprint         = "1709.08222",
  archivePrefix  = "arXiv",
  primaryClass   = "hep-lat"
}

@article{Doring:2018xxx,
  author         = "Döring, M. and Hammer, H.-W. and Mai, M. and Pang,
                    J.-Y. and Rusetsky, A. and Wu, J.",
  title          = "{Three-body spectrum in a finite volume: the role of
                    cubic symmetry}",
  journal        = "Phys. Rev. D",
  volume         = "97",
  year           = "2018",
  number         = "11",
  pages          = "114508",
  doi            = "10.1103/PhysRevD.97.114508",
  eprint         = "1802.03362",
  archivePrefix  = "arXiv",
  primaryClass   = "hep-lat"
}

@article{Yaron:2022rmb,
    author = {Yaron, Roee and Bazak, Betzalel and Sch\"afer, Martin and Barnea,
Nir},
    title = "{Spectrum of light nuclei in a finite volume}",
    eprint = "2206.04497",
    archivePrefix = "arXiv",
    primaryClass = "nucl-th",
    doi = "10.1103/PhysRevD.106.014511",
    journal = "Phys. Rev. D",
    volume = "106",
    number = "1",
    pages = "014511",
    year = "2022"
}

@article{Guo:2021lhz,
  title = {Charged Particles Interaction in Both a Finite Volume and a Uniform
Magnetic Field},
  author = {Guo, Peng and Gasparian, Vladimir},
  year = {2021},
  month = may,
  journal = {Phys. Rev. D},
  volume = {103},
  number = {9},
  pages = {094520},
  publisher = {{American Physical Society}},
  doi = {10.1103/PhysRevD.103.094520}
}

@article{Guo:2021qfu,
  title = {Coulomb Corrections to Two-Particle Interactions in Artificial
Traps},
  author = {Guo, Peng},
  year = {2021},
  month = jun,
  journal = {Phys. Rev. C},
  volume = {103},
  number = {6},
  pages = {064611},
  publisher = {{American Physical Society}},
  doi = {10.1103/PhysRevC.103.064611}
}

@article{Lee:2008fa,
    author = "Lee, Dean",
    title = "{Lattice simulations for few- and many-body systems}",
    eprint = "0804.3501",
    archivePrefix = "arXiv",
    primaryClass = "nucl-th",
    doi = "10.1016/j.ppnp.2008.12.001",
    journal = "Prog. Part. Nucl. Phys.",
    volume = "63",
    pages = "117--154",
    year = "2009"
}

@article{Beane:2014qha,
    author = "Beane, Silas R. and Savage, Martin J.",
    title = "{Two-Particle Elastic Scattering in a Finite Volume Including QED}",
    eprint = "1407.4846",
    archivePrefix = "arXiv",
    primaryClass = "hep-lat",
    reportNumber = "NT@UW-14-14, INT-PUB-14-020",
    doi = "10.1103/PhysRevD.90.074511",
    journal = "Phys. Rev. D",
    volume = "90",
    number = "7",
    pages = "074511",
    year = "2014"
}

@article{Yu:2022nzm,
    author = {Yu, Hang and K{\"o}nig, Sebastian and Lee, Dean},
    title = "{Charged-Particle Bound States in Periodic Boxes}",
    eprint = "2212.14379",
    archivePrefix = "arXiv",
    primaryClass = "nucl-th",
    doi = "10.1103/PhysRevLett.131.212502",
    journal = "Phys. Rev. Lett.",
    volume = "131",
    number = "21",
    pages = "212502",
    year = "2023"
}

@article{Lu:2018bat,
    author = "Lu, Bing-Nan and Li, Ning and Elhatisari, Serdar and Lee, Dean and Epelbaum, Evgeny and Mei{\ss}ner, Ulf-G.",
    title = "{Essential elements for nuclear binding}",
    eprint = "1812.10928",
    archivePrefix = "arXiv",
    primaryClass = "nucl-th",
    doi = "10.1016/j.physletb.2019.134863",
    journal = "Phys. Lett. B",
    volume = "797",
    pages = "134863",
    year = "2019"
}

@misc{Harris:2025zsh,
    author = "Harris, E. and others",
    title = "{Quantifying alpha clustering in the ground states of 16-O and 20-Ne}",
    eprint = "2507.17059",
    archivePrefix = "arXiv",
    primaryClass = "nucl-ex",
    month = "7",
    year = "2025"
}

@misc{Rothman:2025uza,
    author = "Rothman, M. and Johnson-Toth, B. and Hagen, G. and Heinz, M. and Papenbrock, T.",
    title = "{NuLattice: Ab initio computations of atomic nuclei on lattices}",
    eprint = "2509.08771",
    archivePrefix = "arXiv",
    primaryClass = "nucl-th",
    month = "9",
    year = "2025"
}

@article{Bubna:2024izx,
    author = {Bubna, Rishabh and Hammer, Hans-Werner and M{\"u}ller, Fabian and Pang, Jin-Yi and Rusetsky, Akaki and Wu, Jia-Jun},
    title = {{L{\"u}scher equation with long-range forces}},
    eprint = "2402.12985",
    archivePrefix = "arXiv",
    primaryClass = "hep-lat",
    doi = "10.1007/JHEP05(2024)168",
    journal = "JHEP",
    volume = "05",
    pages = "168",
    year = "2024"
}

@article{Zhang:2020rhz,
    author = "Zhang, Xilin and Stroberg, S. R. and Navr{\'a}til, P. and Gwak, Chan and Melendez, J. A. and Furnstahl, R. J. and Holt, J. D.",
    title = "{Ab Initio Calculations of Low-Energy Nuclear Scattering Using Confining Potential Traps}",
    eprint = "2004.13575",
    archivePrefix = "arXiv",
    primaryClass = "nucl-th",
    reportNumber = "INT-PUB-20-018",
    doi = "10.1103/PhysRevLett.125.112503",
    journal = "Phys. Rev. Lett.",
    volume = "125",
    number = "11",
    pages = "112503",
    year = "2020"
}

@article{Luu:2010hw,
    author = "Luu, Thomas and Savage, Martin J. and Schwenk, Achim and Vary, James P.",
    title = "{Nucleon-Nucleon Scattering in a Harmonic Potential}",
    eprint = "1006.0427",
    archivePrefix = "arXiv",
    primaryClass = "nucl-th",
    reportNumber = "NT-UW-10-14",
    doi = "10.1103/PhysRevC.82.034003",
    journal = "Phys. Rev. C",
    volume = "82",
    pages = "034003",
    year = "2010"
}

@article{Bour:2011ef,
    author = "Bour, Shahin and Koenig, Sebastian and Lee, Dean and Hammer, H. -W. and Meissner, Ulf-G.",
    title = "{Topological phases for bound states moving in a finite volume}",
    eprint = "1107.1272",
    archivePrefix = "arXiv",
    primaryClass = "nucl-th",
    doi = "10.1103/PhysRevD.84.091503",
    journal = "Phys. Rev. D",
    volume = "84",
    pages = "091503",
    year = "2011"
}

@article{Briceno:2013hya,
    author = "Briceno, Raul A. and Davoudi, Zohreh and Luu, Thomas C. and Savage, Martin J.",
    title = "{Two-Baryon Systems with Twisted Boundary Conditions}",
    eprint = "1311.7686",
    archivePrefix = "arXiv",
    primaryClass = "hep-lat",
    reportNumber = "INT-PUB-13-045, JLAB-THY-13-1827",
    doi = "10.1103/PhysRevD.89.074509",
    journal = "Phys. Rev. D",
    volume = "89",
    number = "7",
    pages = "074509",
    year = "2014"
}

@misc{Niu:2025uxk,
    author = "Niu, Zhong-Wang and Lu, Bing-Nan",
    title = "{Sign-Problem-Free Nuclear Quantum Monte Carlo}",
    eprint = "2506.12874",
    archivePrefix = "arXiv",
    primaryClass = "nucl-th",
    month = "6",
    year = "2025"
}

@article{deBoer:2017ldl,
    author = "deBoer, R. J. and others",
    title = "{The $^{12}$C({\ensuremath{\alpha}},{\ensuremath{\gamma}})$^{16}$O reaction and its implications for stellar helium burning}",
    eprint = "1709.03144",
    archivePrefix = "arXiv",
    primaryClass = "nucl-ex",
    doi = "10.1103/RevModPhys.89.035007",
    journal = "Rev. Mod. Phys.",
    volume = "89",
    number = "3",
    pages = "035007",
    year = "2017"
}

@article{Barrett:2013nh,
    author = "Barrett, Bruce R. and Navratil, Petr and Vary, James P.",
    title = "{Ab initio no core shell model}",
    doi = "10.1016/j.ppnp.2012.10.003",
    journal = "Prog. Part. Nucl. Phys.",
    volume = "69",
    pages = "131--181",
    year = "2013"
}

@article{Yu:2023ucq,
    author = {Yu, Hang and Yapa, Nuwan and K{\"o}nig, Sebastian},
    title = "{Complex scaling in finite volume}",
    eprint = "2309.03196",
    archivePrefix = "arXiv",
    primaryClass = "nucl-th",
    doi = "10.1103/PhysRevC.109.014316",
    journal = "Phys. Rev. C",
    volume = "109",
    number = "1",
    pages = "014316",
    year = "2024"
}

@article{Busch:1998cey,
    author = "Busch, Thomas and Englert, Berthold-Georg and Rza{\.z}ewski, Kazimierz and Wilkens, Martin",
    title = "{Two Cold Atoms in a Harmonic Trap}",
    doi = "10.1023/a:1018705520999",
    journal = "Found. Phys.",
    volume = "28",
    number = "4",
    pages = "549--559",
    year = "1998"
}

@article{Konig:2022cya,
    author = {K{\"o}nig, S.},
    title = "{Efficient few-body calculations in finite volume}",
    eprint = "2211.00395",
    archivePrefix = "arXiv",
    primaryClass = "nucl-th",
    doi = "10.1088/1742-6596/2453/1/012025",
    journal = "J. Phys. Conf. Ser.",
    volume = "2453",
    number = "1",
    pages = "012025",
    year = "2023"
}

@article{Freer:2017gip,
    author = "Freer, Martin and Horiuchi, Hisashi and Kanada-En'yo, Yoshiko and Lee, Dean and Mei{\ss}ner, Ulf-G.",
    title = "{Microscopic Clustering in Light Nuclei}",
    eprint = "1705.06192",
    archivePrefix = "arXiv",
    primaryClass = "nucl-th",
    doi = "10.1103/RevModPhys.90.035004",
    journal = "Rev. Mod. Phys.",
    volume = "90",
    number = "3",
    pages = "035004",
    year = "2018"
}

@article{Miyagi:2023zvv,
    author = "Miyagi, T. and Cao, X. and Seutin, R. and Bacca, S. and Garcia Ruiz, R. F. and Hebeler, K. and Holt, J. D. and Schwenk, A.",
    title = "{Impact of Two-Body Currents on Magnetic Dipole Moments of Nuclei}",
    eprint = "2311.14383",
    archivePrefix = "arXiv",
    primaryClass = "nucl-th",
    doi = "10.1103/PhysRevLett.132.232503",
    journal = "Phys. Rev. Lett.",
    volume = "132",
    number = "23",
    pages = "232503",
    year = "2024"
}

@article{BMW:2014pzb,
    author = "Borsanyi, Sz. and others",
    collaboration = "BMW",
    title = "{Ab initio calculation of the neutron-proton mass difference}",
    eprint = "1406.4088",
    archivePrefix = "arXiv",
    primaryClass = "hep-lat",
    doi = "10.1126/science.1257050",
    journal = "Science",
    volume = "347",
    pages = "1452--1455",
    year = "2015"
}

@misc{Elhatisari:2025fyu,
    author = "Elhatisari, Serdar and Hildenbrand, Fabian and Mei{\ss}ner, Ulf-G.",
    title = "{Ab initio lattice study of neutron-alpha scattering with chiral forces at N3LO}",
    eprint = "2507.08495",
    archivePrefix = "arXiv",
    primaryClass = "nucl-th",
    month = "7",
    year = "2025"
}

@article{Nollett:2011qf,
    author = "Nollett, Kenneth M. and Wiringa, R. B.",
    title = "{Asymptotic normalization coefficients from ab initio calculations}",
    eprint = "1102.1787",
    archivePrefix = "arXiv",
    primaryClass = "nucl-th",
    doi = "10.1103/PhysRevC.83.041001",
    journal = "Phys. Rev. C",
    volume = "83",
    pages = "041001",
    year = "2011"
}

@article{Hammer:2017tjm,
    author = "Hammer, H.-W. and Ji, C. and Phillips, D. R.",
    title = "{Effective field theory description of halo nuclei}",
    eprint = "1702.08605",
    archivePrefix = "arXiv",
    primaryClass = "nucl-th",
    doi = "10.1088/1361-6471/aa83db",
    journal = "J. Phys. G",
    volume = "44",
    number = "10",
    pages = "103002",
    year = "2017"
}

@article{Entem:2003ft,
    author = "Entem, D. R. and Machleidt, R.",
    title = "{Accurate charge dependent nucleon nucleon potential at fourth order of chiral perturbation theory}",
    eprint = "nucl-th/0304018",
    archivePrefix = "arXiv",
    doi = "10.1103/PhysRevC.68.041001",
    journal = "Phys. Rev. C",
    volume = "68",
    pages = "041001",
    year = "2003"
}
